\documentclass[twocolumn,traditabstract,longauth]{aa}

\def\setsymbol#1#2{\expandafter\def\csname #1\endcsname{#2}}
\def\getsymbol#1{\csname #1\endcsname}

\def\Planck{\textit{Planck}}

\def\all2013resultspapers{\nocite{planck2013-p01, planck2013-p02, planck2013-p02a, planck2013-p02d, planck2013-p02b, planck2013-p03, planck2013-p03c, planck2013-p03f, planck2013-p03d, planck2013-p03e, planck2013-p01a, planck2013-p06, planck2013-p03a, planck2013-pip88, planck2013-p08, planck2013-p11, planck2013-p12, planck2013-p13, planck2013-p14, planck2013-p15, planck2013-p05b, planck2013-p17, planck2013-p09, planck2013-p09a, planck2013-p20, planck2013-p19, planck2013-pipaberration, planck2013-p05, planck2013-p05a, planck2013-pip56, planck2013-p06b}}

\newbox\tablebox    \newdimen\tablewidth
\def\leaderfil{\leaders\hbox to 5pt{\hss.\hss}\hfil}
\def\endPlancktablewide{\tablewidth=\textwidth 
    $$\hss\copy\tablebox\hss$$
    \vskip-\lastskip\vskip -2pt}
\def\tablenote#1 #2\par{\begingroup \parindent=0.8em
    \abovedisplayshortskip=0pt\belowdisplayshortskip=0pt
    \noindent
    $$\hss\vbox{\hsize\tablewidth \hangindent=\parindent \hangafter=1 \noindent
    \hbox to \parindent{$^#1$\hss}\strut#2\strut\par}\hss$$
    \endgroup}
\def\doubleline{\vskip 3pt\hrule \vskip 1.5pt \hrule \vskip 5pt}

\def\L2{\ifmmode L_2\else $L_2$\fi}

\def\DeltaT{\ifmmode \Delta T\else $\Delta T$\fi}
\def\deltat{\ifmmode \Delta t\else $\Delta t$\fi}
\def\fknee{\ifmmode f_{\rm knee}\else $f_{\rm knee}$\fi}
\def\Fmax{\ifmmode F_{\rm max}\else $F_{\rm max}$\fi}
\def\solar{\ifmmode{\rm M}_{\mathord\odot}\else${\rm M}_{\mathord\odot}$\fi}
\def\Msolar{\ifmmode{\rm M}_{\mathord\odot}\else${\rm M}_{\mathord\odot}$\fi}
\def\Lsolar{\ifmmode{\rm L}_{\mathord\odot}\else${\rm L}_{\mathord\odot}$\fi}

\def\inv{\ifmmode^{-1}\else$^{-1}$\fi}
\def\mo{\ifmmode^{-1}\else$^{-1}$\fi}
\def\sup#1{\ifmmode ^{\rm #1}\else $^{\rm #1}$\fi}
\def\expo#1{\ifmmode \times 10^{#1}\else $\times 10^{#1}$\fi}
\def\,{\thinspace}
\def\lsim{\mathrel{\raise .4ex\hbox{\rlap{$<$}\lower 1.2ex\hbox{$\sim$}}}}
\def\gsim{\mathrel{\raise .4ex\hbox{\rlap{$>$}\lower 1.2ex\hbox{$\sim$}}}}

\def\simprop{\mathrel{\raise .4ex\hbox{\rlap{$\propto$}\lower 1.2ex\hbox{$\sim$}}}}
\def\deg{\ifmmode^\circ\else$^\circ$\fi}
\def\pdeg{\ifmmode $\setbox0=\hbox{$^{\circ}$}\rlap{\hskip.11\wd0 .}$^{\circ}
          \else \setbox0=\hbox{$^{\circ}$}\rlap{\hskip.11\wd0 .}$^{\circ}$\fi}
\def\arcs{\ifmmode {^{\scriptstyle\prime\prime}}
          \else $^{\scriptstyle\prime\prime}$\fi}
\def\arcm{\ifmmode {^{\scriptstyle\prime}}
          \else $^{\scriptstyle\prime}$\fi}
\newdimen\sa  \newdimen\sb
\def\parcs{\sa=.07em \sb=.03em
     \ifmmode \hbox{\rlap{.}}^{\scriptstyle\prime\kern -\sb\prime}\hbox{\kern -\sa}
     \else \rlap{.}$^{\scriptstyle\prime\kern -\sb\prime}$\kern -\sa\fi}
\def\parcm{\sa=.08em \sb=.03em
     \ifmmode \hbox{\rlap{.}\kern\sa}^{\scriptstyle\prime}\hbox{\kern-\sb}
     \else \rlap{.}\kern\sa$^{\scriptstyle\prime}$\kern-\sb\fi}
\def\ra[#1 #2 #3.#4]{#1\sup{h}#2\sup{m}#3\sup{s}\llap.#4}
\def\dec[#1 #2 #3.#4]{#1\deg#2\arcm#3\arcs\llap.#4}
\def\deco[#1 #2 #3]{#1\deg#2\arcm#3\arcs}
\def\rra[#1 #2]{#1\sup{h}#2\sup{m}}

\def\dots{\relax\ifmmode \ldots\else $\ldots$\fi}
\def\WHzsr{\ifmmode $W\,Hz\mo\,sr\mo$\else W\,Hz\mo\,sr\mo\fi}
\def\mHz{\ifmmode $\,mHz$\else \,mHz\fi}
\def\GHz{\ifmmode $\,GHz$\else \,GHz\fi}
\def\mKs{\ifmmode $\,mK\,s$^{1/2}\else \,mK\,s$^{1/2}$\fi}
\def\muKs{\ifmmode \,\mu$K\,s$^{1/2}\else \,$\mu$K\,s$^{1/2}$\fi}
\def\muKRJs{\ifmmode \,\mu$K$_{\rm RJ}$\,s$^{1/2}\else \,$\mu$K$_{\rm RJ}$\,s$^{1/2}$\fi}
\def\muKHz{\ifmmode \,\mu$K\,Hz$^{-1/2}\else \,$\mu$K\,Hz$^{-1/2}$\fi}
\def\MJysr{\ifmmode \,$MJy\,sr\mo$\else \,MJy\,sr\mo\fi}
\def\MJysrmK{\ifmmode \,$MJy\,sr\mo$\,mK$_{\rm CMB}\mo\else \,MJy\,sr\mo\,mK$_{\rm CMB}\mo$\fi}
\def\microns{\ifmmode \,\mu$m$\else \,$\mu$m\fi}

\def\muK{\ifmmode \,\mu$K$\else \,$\mu$\hbox{K}\fi}
\def\microK{\ifmmode \,\mu$K$\else \,$\mu$\hbox{K}\fi}
\def\muW{\ifmmode \,\mu$W$\else \,$\mu$\hbox{W}\fi}
\def\kms{\ifmmode $\,km\,s$^{-1}\else \,km\,s$^{-1}$\fi}
\def\kmsMpc{\ifmmode $\,\kms\,Mpc\mo$\else \,\kms\,Mpc\mo\fi}
\setsymbol{LFI:center:frequency:70GHz:units}{70.3\,GHz}
\setsymbol{LFI:center:frequency:44GHz:units}{44.1\,GHz}
\setsymbol{LFI:center:frequency:30GHz:units}{28.5\,GHz}

\setsymbol{LFI:center:frequency:70GHz}{70.3}
\setsymbol{LFI:center:frequency:44GHz}{44.1}
\setsymbol{LFI:center:frequency:30GHz}{28.5}

\setsymbol{LFI:center:frequency:LFI18:Rad:M:units}{71.7\GHz}
\setsymbol{LFI:center:frequency:LFI19:Rad:M:units}{67.5\GHz}
\setsymbol{LFI:center:frequency:LFI20:Rad:M:units}{69.2\GHz}
\setsymbol{LFI:center:frequency:LFI21:Rad:M:units}{70.4\GHz}
\setsymbol{LFI:center:frequency:LFI22:Rad:M:units}{71.5\GHz}
\setsymbol{LFI:center:frequency:LFI23:Rad:M:units}{70.8\GHz}
\setsymbol{LFI:center:frequency:LFI24:Rad:M:units}{44.4\GHz}
\setsymbol{LFI:center:frequency:LFI25:Rad:M:units}{44.0\GHz}
\setsymbol{LFI:center:frequency:LFI26:Rad:M:units}{43.9\GHz}
\setsymbol{LFI:center:frequency:LFI27:Rad:M:units}{28.3\GHz}
\setsymbol{LFI:center:frequency:LFI28:Rad:M:units}{28.8\GHz}
\setsymbol{LFI:center:frequency:LFI18:Rad:S:units}{70.1\GHz}
\setsymbol{LFI:center:frequency:LFI19:Rad:S:units}{69.6\GHz}
\setsymbol{LFI:center:frequency:LFI20:Rad:S:units}{69.5\GHz}
\setsymbol{LFI:center:frequency:LFI21:Rad:S:units}{69.5\GHz}
\setsymbol{LFI:center:frequency:LFI22:Rad:S:units}{72.8\GHz}
\setsymbol{LFI:center:frequency:LFI23:Rad:S:units}{71.3\GHz}
\setsymbol{LFI:center:frequency:LFI24:Rad:S:units}{44.1\GHz}
\setsymbol{LFI:center:frequency:LFI25:Rad:S:units}{44.1\GHz}
\setsymbol{LFI:center:frequency:LFI26:Rad:S:units}{44.1\GHz}
\setsymbol{LFI:center:frequency:LFI27:Rad:S:units}{28.5\GHz}
\setsymbol{LFI:center:frequency:LFI28:Rad:S:units}{28.2\GHz}

\setsymbol{LFI:center:frequency:LFI18:Rad:M}{71.7}
\setsymbol{LFI:center:frequency:LFI19:Rad:M}{67.5}
\setsymbol{LFI:center:frequency:LFI20:Rad:M}{69.2}
\setsymbol{LFI:center:frequency:LFI21:Rad:M}{70.4}
\setsymbol{LFI:center:frequency:LFI22:Rad:M}{71.5}
\setsymbol{LFI:center:frequency:LFI23:Rad:M}{70.8}
\setsymbol{LFI:center:frequency:LFI24:Rad:M}{44.4}
\setsymbol{LFI:center:frequency:LFI25:Rad:M}{44.0}
\setsymbol{LFI:center:frequency:LFI26:Rad:M}{43.9}
\setsymbol{LFI:center:frequency:LFI27:Rad:M}{28.3}
\setsymbol{LFI:center:frequency:LFI28:Rad:M}{28.8}
\setsymbol{LFI:center:frequency:LFI18:Rad:S}{70.1}
\setsymbol{LFI:center:frequency:LFI19:Rad:S}{69.6}
\setsymbol{LFI:center:frequency:LFI20:Rad:S}{69.5}
\setsymbol{LFI:center:frequency:LFI21:Rad:S}{69.5}
\setsymbol{LFI:center:frequency:LFI22:Rad:S}{72.8}
\setsymbol{LFI:center:frequency:LFI23:Rad:S}{71.3}
\setsymbol{LFI:center:frequency:LFI24:Rad:S}{44.1}
\setsymbol{LFI:center:frequency:LFI25:Rad:S}{44.1}
\setsymbol{LFI:center:frequency:LFI26:Rad:S}{44.1}
\setsymbol{LFI:center:frequency:LFI27:Rad:S}{28.5}
\setsymbol{LFI:center:frequency:LFI28:Rad:S}{28.2}

\setsymbol{LFI:white:noise:sensitivity:70GHz:units}{134.7\muKs}
\setsymbol{LFI:white:noise:sensitivity:44GHz:units}{164.7\muKs}
\setsymbol{LFI:white:noise:sensitivity:30GHz:units}{143.4\muKs}

\setsymbol{LFI:white:noise:sensitivity:70GHz}{134.7}
\setsymbol{LFI:white:noise:sensitivity:44GHz}{164.7}
\setsymbol{LFI:white:noise:sensitivity:30GHz}{143.4}

\setsymbol{LFI:white:noise:sensitivity:LFI18:Rad:M:units}{512.0\muKs}
\setsymbol{LFI:white:noise:sensitivity:LFI19:Rad:M:units}{581.4\muKs}
\setsymbol{LFI:white:noise:sensitivity:LFI20:Rad:M:units}{590.8\muKs}
\setsymbol{LFI:white:noise:sensitivity:LFI21:Rad:M:units}{455.2\muKs}
\setsymbol{LFI:white:noise:sensitivity:LFI22:Rad:M:units}{492.0\muKs}
\setsymbol{LFI:white:noise:sensitivity:LFI23:Rad:M:units}{507.7\muKs}
\setsymbol{LFI:white:noise:sensitivity:LFI24:Rad:M:units}{462.2\muKs}
\setsymbol{LFI:white:noise:sensitivity:LFI25:Rad:M:units}{413.6\muKs}
\setsymbol{LFI:white:noise:sensitivity:LFI26:Rad:M:units}{478.6\muKs}
\setsymbol{LFI:white:noise:sensitivity:LFI27:Rad:M:units}{277.7\muKs}
\setsymbol{LFI:white:noise:sensitivity:LFI28:Rad:M:units}{312.3\muKs}
\setsymbol{LFI:white:noise:sensitivity:LFI18:Rad:S:units}{465.7\muKs}
\setsymbol{LFI:white:noise:sensitivity:LFI19:Rad:S:units}{555.6\muKs}
\setsymbol{LFI:white:noise:sensitivity:LFI20:Rad:S:units}{623.2\muKs}
\setsymbol{LFI:white:noise:sensitivity:LFI21:Rad:S:units}{564.1\muKs}
\setsymbol{LFI:white:noise:sensitivity:LFI22:Rad:S:units}{534.4\muKs}
\setsymbol{LFI:white:noise:sensitivity:LFI23:Rad:S:units}{542.4\muKs}
\setsymbol{LFI:white:noise:sensitivity:LFI24:Rad:S:units}{399.2\muKs}
\setsymbol{LFI:white:noise:sensitivity:LFI25:Rad:S:units}{392.6\muKs}
\setsymbol{LFI:white:noise:sensitivity:LFI26:Rad:S:units}{418.6\muKs}
\setsymbol{LFI:white:noise:sensitivity:LFI27:Rad:S:units}{302.9\muKs}
\setsymbol{LFI:white:noise:sensitivity:LFI28:Rad:S:units}{285.3\muKs}

\setsymbol{LFI:white:noise:sensitivity:LFI18:Rad:M}{512.0}
\setsymbol{LFI:white:noise:sensitivity:LFI19:Rad:M}{581.4}
\setsymbol{LFI:white:noise:sensitivity:LFI20:Rad:M}{590.8}
\setsymbol{LFI:white:noise:sensitivity:LFI21:Rad:M}{455.2}
\setsymbol{LFI:white:noise:sensitivity:LFI22:Rad:M}{492.0}
\setsymbol{LFI:white:noise:sensitivity:LFI23:Rad:M}{507.7}
\setsymbol{LFI:white:noise:sensitivity:LFI24:Rad:M}{462.2}
\setsymbol{LFI:white:noise:sensitivity:LFI25:Rad:M}{413.6}
\setsymbol{LFI:white:noise:sensitivity:LFI26:Rad:M}{478.6}
\setsymbol{LFI:white:noise:sensitivity:LFI27:Rad:M}{277.7}
\setsymbol{LFI:white:noise:sensitivity:LFI28:Rad:M}{312.3}
\setsymbol{LFI:white:noise:sensitivity:LFI18:Rad:S}{465.7}
\setsymbol{LFI:white:noise:sensitivity:LFI19:Rad:S}{555.6}
\setsymbol{LFI:white:noise:sensitivity:LFI20:Rad:S}{623.2}
\setsymbol{LFI:white:noise:sensitivity:LFI21:Rad:S}{564.1}
\setsymbol{LFI:white:noise:sensitivity:LFI22:Rad:S}{534.4}
\setsymbol{LFI:white:noise:sensitivity:LFI23:Rad:S}{542.4}
\setsymbol{LFI:white:noise:sensitivity:LFI24:Rad:S}{399.2}
\setsymbol{LFI:white:noise:sensitivity:LFI25:Rad:S}{392.6}
\setsymbol{LFI:white:noise:sensitivity:LFI26:Rad:S}{418.6}
\setsymbol{LFI:white:noise:sensitivity:LFI27:Rad:S}{302.9}
\setsymbol{LFI:white:noise:sensitivity:LFI28:Rad:S}{285.3}

\setsymbol{LFI:knee:frequency:70GHz:units}{29.5\mHz}
\setsymbol{LFI:knee:frequency:44GHz:units}{56.2\mHz}
\setsymbol{LFI:knee:frequency:30GHz:units}{113.7\mHz}

\setsymbol{LFI:knee:frequency:70GHz}{29.5}
\setsymbol{LFI:knee:frequency:44GHz}{56.2}
\setsymbol{LFI:knee:frequency:30GHz}{113.7}

\setsymbol{LFI:knee:frequency:LFI18:Rad:M:units}{16.3\mHz}
\setsymbol{LFI:knee:frequency:LFI19:Rad:M:units}{15.1\mHz}
\setsymbol{LFI:knee:frequency:LFI20:Rad:M:units}{18.7\mHz}
\setsymbol{LFI:knee:frequency:LFI21:Rad:M:units}{37.2\mHz}
\setsymbol{LFI:knee:frequency:LFI22:Rad:M:units}{12.7\mHz}
\setsymbol{LFI:knee:frequency:LFI23:Rad:M:units}{34.6\mHz}
\setsymbol{LFI:knee:frequency:LFI24:Rad:M:units}{46.2\mHz}
\setsymbol{LFI:knee:frequency:LFI25:Rad:M:units}{24.9\mHz}
\setsymbol{LFI:knee:frequency:LFI26:Rad:M:units}{67.6\mHz}
\setsymbol{LFI:knee:frequency:LFI27:Rad:M:units}{187.4\mHz}
\setsymbol{LFI:knee:frequency:LFI28:Rad:M:units}{122.2\mHz}
\setsymbol{LFI:knee:frequency:LFI18:Rad:S:units}{17.7\mHz}
\setsymbol{LFI:knee:frequency:LFI19:Rad:S:units}{22.0\mHz}
\setsymbol{LFI:knee:frequency:LFI20:Rad:S:units}{8.7\mHz}
\setsymbol{LFI:knee:frequency:LFI21:Rad:S:units}{25.9\mHz}
\setsymbol{LFI:knee:frequency:LFI22:Rad:S:units}{15.8\mHz}
\setsymbol{LFI:knee:frequency:LFI23:Rad:S:units}{129.8\mHz}
\setsymbol{LFI:knee:frequency:LFI24:Rad:S:units}{100.9\mHz}
\setsymbol{LFI:knee:frequency:LFI25:Rad:S:units}{38.9\mHz}
\setsymbol{LFI:knee:frequency:LFI26:Rad:S:units}{58.9\mHz}
\setsymbol{LFI:knee:frequency:LFI27:Rad:S:units}{104.4\mHz}
\setsymbol{LFI:knee:frequency:LFI28:Rad:S:units}{40.7\mHz}

\setsymbol{LFI:knee:frequency:LFI18:Rad:M}{16.3}
\setsymbol{LFI:knee:frequency:LFI19:Rad:M}{15.1}
\setsymbol{LFI:knee:frequency:LFI20:Rad:M}{18.7}
\setsymbol{LFI:knee:frequency:LFI21:Rad:M}{37.2}
\setsymbol{LFI:knee:frequency:LFI22:Rad:M}{12.7}
\setsymbol{LFI:knee:frequency:LFI23:Rad:M}{34.6}
\setsymbol{LFI:knee:frequency:LFI24:Rad:M}{46.2}
\setsymbol{LFI:knee:frequency:LFI25:Rad:M}{24.9}
\setsymbol{LFI:knee:frequency:LFI26:Rad:M}{67.6}
\setsymbol{LFI:knee:frequency:LFI27:Rad:M}{187.4}
\setsymbol{LFI:knee:frequency:LFI28:Rad:M}{122.2}
\setsymbol{LFI:knee:frequency:LFI18:Rad:S}{17.7}
\setsymbol{LFI:knee:frequency:LFI19:Rad:S}{22.0}
\setsymbol{LFI:knee:frequency:LFI20:Rad:S}{8.7}
\setsymbol{LFI:knee:frequency:LFI21:Rad:S}{25.9}
\setsymbol{LFI:knee:frequency:LFI22:Rad:S}{15.8}
\setsymbol{LFI:knee:frequency:LFI23:Rad:S}{129.8}
\setsymbol{LFI:knee:frequency:LFI24:Rad:S}{100.9}
\setsymbol{LFI:knee:frequency:LFI25:Rad:S}{38.9}
\setsymbol{LFI:knee:frequency:LFI26:Rad:S}{58.9}
\setsymbol{LFI:knee:frequency:LFI27:Rad:S}{104.4}
\setsymbol{LFI:knee:frequency:LFI28:Rad:S}{40.7}

\setsymbol{LFI:slope:70GHz:units}{$-1.03$\mHz}
\setsymbol{LFI:slope:44GHz:units}{$-0.89$\mHz}
\setsymbol{LFI:slope:30GHz:units}{$-0.87$\mHz}

\setsymbol{LFI:slope:70GHz}{$-1.03$}
\setsymbol{LFI:slope:44GHz}{$-0.89$}
\setsymbol{LFI:slope:30GHz}{$-0.87$}

\setsymbol{LFI:slope:LFI18:Rad:M:units}{$-1.04$\mHz}
\setsymbol{LFI:slope:LFI19:Rad:M:units}{$-1.09$\mHz}
\setsymbol{LFI:slope:LFI20:Rad:M:units}{$-0.69$\mHz}
\setsymbol{LFI:slope:LFI21:Rad:M:units}{$-1.56$\mHz}
\setsymbol{LFI:slope:LFI22:Rad:M:units}{$-1.01$\mHz}
\setsymbol{LFI:slope:LFI23:Rad:M:units}{$-0.96$\mHz}
\setsymbol{LFI:slope:LFI24:Rad:M:units}{$-0.83$\mHz}
\setsymbol{LFI:slope:LFI25:Rad:M:units}{$-0.91$\mHz}
\setsymbol{LFI:slope:LFI26:Rad:M:units}{$-0.95$\mHz}
\setsymbol{LFI:slope:LFI27:Rad:M:units}{$-0.87$\mHz}
\setsymbol{LFI:slope:LFI28:Rad:M:units}{$-0.88$\mHz}
\setsymbol{LFI:slope:LFI18:Rad:S:units}{$-1.15$\mHz}
\setsymbol{LFI:slope:LFI19:Rad:S:units}{$-1.00$\mHz}
\setsymbol{LFI:slope:LFI20:Rad:S:units}{$-0.95$\mHz}
\setsymbol{LFI:slope:LFI21:Rad:S:units}{$-0.92$\mHz}
\setsymbol{LFI:slope:LFI22:Rad:S:units}{$-1.01$\mHz}
\setsymbol{LFI:slope:LFI23:Rad:S:units}{$-0.95$\mHz}
\setsymbol{LFI:slope:LFI24:Rad:S:units}{$-0.73$\mHz}
\setsymbol{LFI:slope:LFI25:Rad:S:units}{$-1.16$\mHz}
\setsymbol{LFI:slope:LFI26:Rad:S:units}{$-0.79$\mHz}
\setsymbol{LFI:slope:LFI27:Rad:S:units}{$-0.82$\mHz}
\setsymbol{LFI:slope:LFI28:Rad:S:units}{$-0.91$\mHz}

\setsymbol{LFI:slope:LFI18:Rad:M}{$-1.04$}
\setsymbol{LFI:slope:LFI19:Rad:M}{$-1.09$}
\setsymbol{LFI:slope:LFI20:Rad:M}{$-0.69$}
\setsymbol{LFI:slope:LFI21:Rad:M}{$-1.56$}
\setsymbol{LFI:slope:LFI22:Rad:M}{$-1.01$}
\setsymbol{LFI:slope:LFI23:Rad:M}{$-0.96$}
\setsymbol{LFI:slope:LFI24:Rad:M}{$-0.83$}
\setsymbol{LFI:slope:LFI25:Rad:M}{$-0.91$}
\setsymbol{LFI:slope:LFI26:Rad:M}{$-0.95$}
\setsymbol{LFI:slope:LFI27:Rad:M}{$-0.87$}
\setsymbol{LFI:slope:LFI28:Rad:M}{$-0.88$}
\setsymbol{LFI:slope:LFI18:Rad:S}{$-1.15$}
\setsymbol{LFI:slope:LFI19:Rad:S}{$-1.00$}
\setsymbol{LFI:slope:LFI20:Rad:S}{$-0.95$}
\setsymbol{LFI:slope:LFI21:Rad:S}{$-0.92$}
\setsymbol{LFI:slope:LFI22:Rad:S}{$-1.01$}
\setsymbol{LFI:slope:LFI23:Rad:S}{$-0.95$}
\setsymbol{LFI:slope:LFI24:Rad:S}{$-0.73$}
\setsymbol{LFI:slope:LFI25:Rad:S}{$-1.16$}
\setsymbol{LFI:slope:LFI26:Rad:S}{$-0.79$}
\setsymbol{LFI:slope:LFI27:Rad:S}{$-0.82$}
\setsymbol{LFI:slope:LFI28:Rad:S}{$-0.91$}

\setsymbol{LFI:FWHM:70GHz:units}{13\parcm01}
\setsymbol{LFI:FWHM:44GHz:units}{27\parcm92}
\setsymbol{LFI:FWHM:30GHz:units}{32\parcm65}

\setsymbol{LFI:FWHM:70GHz}{13.01}
\setsymbol{LFI:FWHM:44GHz}{27.92}
\setsymbol{LFI:FWHM:30GHz}{32.65}

\setsymbol{LFI:FWHM:LFI18:units}{13\parcm39}
\setsymbol{LFI:FWHM:LFI19:units}{13\parcm01}
\setsymbol{LFI:FWHM:LFI20:units}{12\parcm75}
\setsymbol{LFI:FWHM:LFI21:units}{12\parcm74}
\setsymbol{LFI:FWHM:LFI22:units}{12\parcm87}
\setsymbol{LFI:FWHM:LFI23:units}{13\parcm27}
\setsymbol{LFI:FWHM:LFI24:units}{22\parcm98}
\setsymbol{LFI:FWHM:LFI25:units}{30\parcm46}
\setsymbol{LFI:FWHM:LFI26:units}{30\parcm31}
\setsymbol{LFI:FWHM:LFI27:units}{32\parcm65}
\setsymbol{LFI:FWHM:LFI28:units}{32\parcm66}

\setsymbol{LFI:FWHM:LFI18}{13.39}
\setsymbol{LFI:FWHM:LFI19}{13.01}
\setsymbol{LFI:FWHM:LFI20}{12.75}
\setsymbol{LFI:FWHM:LFI21}{12.74}
\setsymbol{LFI:FWHM:LFI22}{12.87}
\setsymbol{LFI:FWHM:LFI23}{13.27}
\setsymbol{LFI:FWHM:LFI24}{22.98}
\setsymbol{LFI:FWHM:LFI25}{30.46}
\setsymbol{LFI:FWHM:LFI26}{30.31}
\setsymbol{LFI:FWHM:LFI27}{32.65}
\setsymbol{LFI:FWHM:LFI28}{32.66}

\setsymbol{LFI:FWHM:uncertainty:LFI18:units}{0.170\arcm}
\setsymbol{LFI:FWHM:uncertainty:LFI19:units}{0.174\arcm}
\setsymbol{LFI:FWHM:uncertainty:LFI20:units}{0.170\arcm}
\setsymbol{LFI:FWHM:uncertainty:LFI21:units}{0.156\arcm}
\setsymbol{LFI:FWHM:uncertainty:LFI22:units}{0.164\arcm}
\setsymbol{LFI:FWHM:uncertainty:LFI23:units}{0.171\arcm}
\setsymbol{LFI:FWHM:uncertainty:LFI24:units}{0.652\arcm}
\setsymbol{LFI:FWHM:uncertainty:LFI25:units}{1.075\arcm}
\setsymbol{LFI:FWHM:uncertainty:LFI26:units}{1.131\arcm}
\setsymbol{LFI:FWHM:uncertainty:LFI27:units}{1.266\arcm}
\setsymbol{LFI:FWHM:uncertainty:LFI28:units}{1.287\arcm}

\setsymbol{LFI:FWHM:uncertainty:LFI18}{0.170}
\setsymbol{LFI:FWHM:uncertainty:LFI19}{0.174}
\setsymbol{LFI:FWHM:uncertainty:LFI20}{0.170}
\setsymbol{LFI:FWHM:uncertainty:LFI21}{0.156}
\setsymbol{LFI:FWHM:uncertainty:LFI22}{0.164}
\setsymbol{LFI:FWHM:uncertainty:LFI23}{0.171}
\setsymbol{LFI:FWHM:uncertainty:LFI24}{0.652}
\setsymbol{LFI:FWHM:uncertainty:LFI25}{1.075}
\setsymbol{LFI:FWHM:uncertainty:LFI26}{1.131}
\setsymbol{LFI:FWHM:uncertainty:LFI27}{1.266}
\setsymbol{LFI:FWHM:uncertainty:LFI28}{1.287}

\setsymbol{HFI:center:frequency:100GHz:units}{100\,GHz}
\setsymbol{HFI:center:frequency:143GHz:units}{143\,GHz}
\setsymbol{HFI:center:frequency:217GHz:units}{217\,GHz}
\setsymbol{HFI:center:frequency:353GHz:units}{353\,GHz}
\setsymbol{HFI:center:frequency:545GHz:units}{545\,GHz}
\setsymbol{HFI:center:frequency:857GHz:units}{857\,GHz}

\setsymbol{HFI:center:frequency:100GHz}{100}
\setsymbol{HFI:center:frequency:143GHz}{143}
\setsymbol{HFI:center:frequency:217GHz}{217}
\setsymbol{HFI:center:frequency:353GHz}{353}
\setsymbol{HFI:center:frequency:545GHz}{545}
\setsymbol{HFI:center:frequency:857GHz}{857}

\setsymbol{HFI:Ndetectors:100GHz}{8}
\setsymbol{HFI:Ndetectors:143GHz}{11}
\setsymbol{HFI:Ndetectors:217GHz}{12}
\setsymbol{HFI:Ndetectors:353GHz}{12}
\setsymbol{HFI:Ndetectors:545GHz}{3}
\setsymbol{HFI:Ndetectors:857GHz}{4}

\setsymbol{HFI:FWHM:Maps:100GHz:units}{9\parcm88}
\setsymbol{HFI:FWHM:Maps:143GHz:units}{7\parcm18}
\setsymbol{HFI:FWHM:Maps:217GHz:units}{4\parcm87}
\setsymbol{HFI:FWHM:Maps:353GHz:units}{4\parcm65}
\setsymbol{HFI:FWHM:Maps:545GHz:units}{4\parcm72}
\setsymbol{HFI:FWHM:Maps:857GHz:units}{4\parcm39}
\setsymbol{HFI:FWHM:Maps:100GHz}{9.88}
\setsymbol{HFI:FWHM:Maps:143GHz}{7.18}
\setsymbol{HFI:FWHM:Maps:217GHz}{4.87}
\setsymbol{HFI:FWHM:Maps:353GHz}{4.65}
\setsymbol{HFI:FWHM:Maps:545GHz}{4.72}
\setsymbol{HFI:FWHM:Maps:857GHz}{4.39}

\setsymbol{HFI:beam:ellipticity:Maps:100GHz}{1.15}
\setsymbol{HFI:beam:ellipticity:Maps:143GHz}{1.01}
\setsymbol{HFI:beam:ellipticity:Maps:217GHz}{1.06}
\setsymbol{HFI:beam:ellipticity:Maps:353GHz}{1.05}
\setsymbol{HFI:beam:ellipticity:Maps:545GHz}{1.14}
\setsymbol{HFI:beam:ellipticity:Maps:857GHz}{1.19}

\setsymbol{HFI:FWHM:Mars:100GHz:units}{9\parcm37}
\setsymbol{HFI:FWHM:Mars:143GHz:units}{7\parcm04}
\setsymbol{HFI:FWHM:Mars:217GHz:units}{4\parcm68}
\setsymbol{HFI:FWHM:Mars:353GHz:units}{4\parcm43}
\setsymbol{HFI:FWHM:Mars:545GHz:units}{3\parcm80}
\setsymbol{HFI:FWHM:Mars:857GHz:units}{3\parcm67}

\setsymbol{HFI:FWHM:Mars:100GHz}{9.37}
\setsymbol{HFI:FWHM:Mars:143GHz}{7.04}
\setsymbol{HFI:FWHM:Mars:217GHz}{4.68}
\setsymbol{HFI:FWHM:Mars:353GHz}{4.43}
\setsymbol{HFI:FWHM:Mars:545GHz}{3.80}
\setsymbol{HFI:FWHM:Mars:857GHz}{3.67}

\setsymbol{HFI:beam:ellipticity:Mars:100GHz}{1.18}
\setsymbol{HFI:beam:ellipticity:Mars:143GHz}{1.03}
\setsymbol{HFI:beam:ellipticity:Mars:217GHz}{1.14}
\setsymbol{HFI:beam:ellipticity:Mars:353GHz}{1.09}
\setsymbol{HFI:beam:ellipticity:Mars:545GHz}{1.25}
\setsymbol{HFI:beam:ellipticity:Mars:857GHz}{1.03}

\setsymbol{HFI:CMB:relative:calibration:100GHz}{$\lsim 1\%$}
\setsymbol{HFI:CMB:relative:calibration:143GHz}{$\lsim 1\%$}
\setsymbol{HFI:CMB:relative:calibration:217GHz}{$\lsim 1\%$}
\setsymbol{HFI:CMB:relative:calibration:353GHz}{$\lsim 1\%$}
\setsymbol{HFI:CMB:relative:calibration:545GHz}{}
\setsymbol{HFI:CMB:relative:calibration:857GHz}{}

\setsymbol{HFI:CMB:absolute:calibration:100GHz}{$\lsim 2\%$}
\setsymbol{HFI:CMB:absolute:calibration:143GHz}{$\lsim 2\%$}
\setsymbol{HFI:CMB:absolute:calibration:217GHz}{$\lsim 2\%$}
\setsymbol{HFI:CMB:absolute:calibration:353GHz}{$\lsim 2\%$}
\setsymbol{HFI:CMB:absolute:calibration:545GHz}{}
\setsymbol{HFI:CMB:absolute:calibration:857GHz}{}

\setsymbol{HFI:FIRAS:gain:calibration:accuracy:statistical:100GHz}{}
\setsymbol{HFI:FIRAS:gain:calibration:accuracy:statistical:143GHz}{}
\setsymbol{HFI:FIRAS:gain:calibration:accuracy:statistical:217GHz}{}
\setsymbol{HFI:FIRAS:gain:calibration:accuracy:statistical:353GHz}{2.5\%}
\setsymbol{HFI:FIRAS:gain:calibration:accuracy:statistical:545GHz}{1\%}
\setsymbol{HFI:FIRAS:gain:calibration:accuracy:statistical:857GHz}{0.5\%}

\setsymbol{HFI:FIRAS:gain:calibration:accuracy:systematic:100GHz}{}
\setsymbol{HFI:FIRAS:gain:calibration:accuracy:systematic:143GHz}{}
\setsymbol{HFI:FIRAS:gain:calibration:accuracy:systematic:217GHz}{}
\setsymbol{HFI:FIRAS:gain:calibration:accuracy:systematic:353GHz}{}
\setsymbol{HFI:FIRAS:gain:calibration:accuracy:systematic:545GHz}{7\%}
\setsymbol{HFI:FIRAS:gain:calibration:accuracy:systematic:857GHz}{7\%}

\setsymbol{HFI:FIRAS:zero:point:accuracy:100GHz:units}{0.8\MJysr}
\setsymbol{HFI:FIRAS:zero:point:accuracy:143GHz:units}{}
\setsymbol{HFI:FIRAS:zero:point:accuracy:217GHz:units}{}
\setsymbol{HFI:FIRAS:zero:point:accuracy:353GHz:units}{1.4\MJysr}
\setsymbol{HFI:FIRAS:zero:point:accuracy:545GHz:units}{2.2\MJysr}
\setsymbol{HFI:FIRAS:zero:point:accuracy:857GHz:units}{1.7\MJysr}

\setsymbol{HFI:FIRAS:zero:point:accuracy:100GHz}{0.8}
\setsymbol{HFI:FIRAS:zero:point:accuracy:143GHz}{}
\setsymbol{HFI:FIRAS:zero:point:accuracy:217GHz}{}
\setsymbol{HFI:FIRAS:zero:point:accuracy:353GHz}{1.4}
\setsymbol{HFI:FIRAS:zero:point:accuracy:545GHz}{2.2}
\setsymbol{HFI:FIRAS:zero:point:accuracy:857GHz}{1.7}

\setsymbol{HFI:unit:conversion:100GHz:units}{0.2415\MJysrmK}
\setsymbol{HFI:unit:conversion:143GHz:units}{0.3694\MJysrmK}
\setsymbol{HFI:unit:conversion:217GHz:units}{0.4811\MJysrmK}
\setsymbol{HFI:unit:conversion:353GHz:units}{0.2883\MJysrmK}
\setsymbol{HFI:unit:conversion:545GHz:units}{0.05826\MJysrmK}
\setsymbol{HFI:unit:conversion:857GHz:units}{0.002238\MJysrmK}

\setsymbol{HFI:unit:conversion:100GHz}{0.2415}
\setsymbol{HFI:unit:conversion:143GHz}{0.3694}
\setsymbol{HFI:unit:conversion:217GHz}{0.4811}
\setsymbol{HFI:unit:conversion:353GHz}{0.2883}
\setsymbol{HFI:unit:conversion:545GHz}{0.05826}
\setsymbol{HFI:unit:conversion:857GHz}{0.002238}

\setsymbol{HFI:colour:correction:alpha=-2:V1.01:100GHz}{0.9893}
\setsymbol{HFI:colour:correction:alpha=-2:V1.01:143GHz}{0.9759}
\setsymbol{HFI:colour:correction:alpha=-2:V1.01:217GHz}{1.0007}
\setsymbol{HFI:colour:correction:alpha=-2:V1.01:353GHz}{1.0028}
\setsymbol{HFI:colour:correction:alpha=-2:V1.01:545GHz}{1.0019}
\setsymbol{HFI:colour:correction:alpha=-2:V1.01:857GHz}{0.9889}

\setsymbol{HFI:colour:correction:alpha=0:V1.01:100GHz}{1.0008}
\setsymbol{HFI:colour:correction:alpha=0:V1.01:143GHz}{1.0148}
\setsymbol{HFI:colour:correction:alpha=0:V1.01:217GHz}{0.9909}
\setsymbol{HFI:colour:correction:alpha=0:V1.01:353GHz}{0.9888}
\setsymbol{HFI:colour:correction:alpha=0:V1.01:545GHz}{0.9878}
\setsymbol{HFI:colour:correction:alpha=0:V1.01:857GHz}{1.0014}

\providecommand{\sorthelp}[1]{}

\usepackage{amsmath}
\usepackage{natbib}
\usepackage[breaklinks,colorlinks,citecolor=blue]{hyperref}
\usepackage{breakurl}
\usepackage{graphicx}
\usepackage{fixltx2e}
\usepackage{url}
\usepackage{txfonts}
\usepackage[switch,pagewise]{lineno}
\newcommand{\ha}{H$\alpha$} 
\newcommand{\hi}{\ion{H}{i}} 
\newcommand{\hii}{\ion{H}{ii}} 
\usepackage{color}
\definecolor{gold}{rgb}{0.85,.66,0}

\begin{document}
\author{\small
Planck Collaboration:
N.~Aghanim\inst{54}
\and
M.~I.~R.~Alves\inst{54, 86, 9}\thanks{Corresponding author: M. I. R. Alves \url{marta.alves@irap.omp.eu}}
\and
M.~Arnaud\inst{66}
\and
D.~Arzoumanian\inst{54}
\and
J.~Aumont\inst{54}
\and
C.~Baccigalupi\inst{79}
\and
A.~J.~Banday\inst{86, 9}
\and
R.~B.~Barreiro\inst{59}
\and
N.~Bartolo\inst{27, 60}
% ...
}
\institute{\small
APC, AstroParticule et Cosmologie, Universit\'{e} Paris Diderot, CNRS/IN2P3, CEA/lrfu, Observatoire de Paris, Sorbonne Paris Cit\'{e}, 10, rue Alice Domon et L\'{e}onie Duquet, 75205 Paris Cedex 13, France\goodbreak
\and
African Institute for Mathematical Sciences, 6-8 Melrose Road, Muizenberg, Cape Town, South Africa\goodbreak
\and
Agenzia Spaziale Italiana Science Data Center, Via del Politecnico snc, 00133, Roma, Italy\goodbreak
\and
Agenzia Spaziale Italiana, Viale Liegi 26, Roma, Italy\goodbreak
\and
Aix Marseille Universit\'{e}, CNRS, LAM (Laboratoire d'Astrophysique de Marseille) UMR 7326, 13388, Marseille, France\goodbreak
\and
Astrophysics Group, Cavendish Laboratory, University of Cambridge, J J Thomson Avenue, Cambridge CB3 0HE, U.K.\goodbreak
\and
Astrophysics \& Cosmology Research Unit, School of Mathematics, Statistics \& Computer Science, University of KwaZulu-Natal, Westville Campus, Private Bag X54001, Durban 4000, South Africa\goodbreak
\and
CITA, University of Toronto, 60 St. George St., Toronto, ON M5S 3H8, Canada\goodbreak
\and
CNRS, IRAP, 9 Av. colonel Roche, BP 44346, F-31028 Toulouse cedex 4, France\goodbreak
\and
CRANN, Trinity College, Dublin, Ireland\goodbreak
\and
California Institute of Technology, Pasadena, California, U.S.A.\goodbreak
\and
Computational Cosmology Center, Lawrence Berkeley National Laboratory, Berkeley, California, U.S.A.\goodbreak
\and
Consejo Superior de Investigaciones Cient\'{\i}ficas (CSIC), Madrid, Spain\goodbreak
\and
DSM/Irfu/SPP, CEA-Saclay, F-91191 Gif-sur-Yvette Cedex, France\goodbreak
\and
DTU Space, National Space Institute, Technical University of Denmark, Elektrovej 327, DK-2800 Kgs. Lyngby, Denmark\goodbreak
\and
D\'{e}partement de Physique Th\'{e}orique, Universit\'{e} de Gen\`{e}ve, 24, Quai E. Ansermet,1211 Gen\`{e}ve 4, Switzerland\goodbreak
\and
Departamento de F\'{\i}sica, Universidad de Oviedo, Avda. Calvo Sotelo s/n, Oviedo, Spain\goodbreak
\and
Department of Astronomy and Astrophysics, University of Toronto, 50 Saint George Street, Toronto, Ontario, Canada\goodbreak
\and
Department of Physics \& Astronomy, University of British Columbia, 6224 Agricultural Road, Vancouver, British Columbia, Canada\goodbreak
\and
Department of Physics and Astronomy, Dana and David Dornsife College of Letter, Arts and Sciences, University of Southern California, Los Angeles, CA 90089, U.S.A.\goodbreak
\and
Department of Physics and Astronomy, University College London, London WC1E 6BT, U.K.\goodbreak
\and
Department of Physics, Florida State University, Keen Physics Building, 77 Chieftan Way, Tallahassee, Florida, U.S.A.\goodbreak
\and
Department of Physics, Gustaf H\"{a}llstr\"{o}min katu 2a, University of Helsinki, Helsinki, Finland\goodbreak
\and
Department of Physics, Princeton University, Princeton, New Jersey, U.S.A.\goodbreak
\and
Department of Physics, University of California, Santa Barbara, California, U.S.A.\goodbreak
\and
Department of Physics, University of Illinois at Urbana-Champaign, 1110 West Green Street, Urbana, Illinois, U.S.A.\goodbreak
\and
Dipartimento di Fisica e Astronomia G. Galilei, Universit\`{a} degli Studi di Padova, via Marzolo 8, 35131 Padova, Italy\goodbreak
\and
Dipartimento di Fisica e Scienze della Terra, Universit\`{a} di Ferrara, Via Saragat 1, 44122 Ferrara, Italy\goodbreak
\and
Dipartimento di Fisica, Universit\`{a} La Sapienza, P. le A. Moro 2, Roma, Italy\goodbreak
\and
Dipartimento di Fisica, Universit\`{a} degli Studi di Milano, Via Celoria, 16, Milano, Italy\goodbreak
\and
Dipartimento di Fisica, Universit\`{a} degli Studi di Trieste, via A. Valerio 2, Trieste, Italy\goodbreak
\and
Dipartimento di Matematica, Universit\`{a} di Roma Tor Vergata, Via della Ricerca Scientifica, 1, Roma, Italy\goodbreak
\and
Discovery Center, Niels Bohr Institute, Blegdamsvej 17, Copenhagen, Denmark\goodbreak
\and
Dpto. Astrof\'{i}sica, Universidad de La Laguna (ULL), E-38206 La Laguna, Tenerife, Spain\goodbreak
\and
European Space Agency, ESAC, Planck Science Office, Camino bajo del Castillo, s/n, Urbanizaci\'{o}n Villafranca del Castillo, Villanueva de la Ca\~{n}ada, Madrid, Spain\goodbreak
\and
European Space Agency, ESTEC, Keplerlaan 1, 2201 AZ Noordwijk, The Netherlands\goodbreak
\and
Facolt\`{a} di Ingegneria, Universit\`{a} degli Studi e-Campus, Via Isimbardi 10, Novedrate (CO), 22060, Italy\goodbreak
\and
Gran Sasso Science Institute, INFN, viale F. Crispi 7, 67100 L'Aquila, Italy\goodbreak
\and
HGSFP and University of Heidelberg, Theoretical Physics Department, Philosophenweg 16, 69120, Heidelberg, Germany\goodbreak
\and
Helsinki Institute of Physics, Gustaf H\"{a}llstr\"{o}min katu 2, University of Helsinki, Helsinki, Finland\goodbreak
\and
INAF - Osservatorio Astronomico di Padova, Vicolo dell'Osservatorio 5, Padova, Italy\goodbreak
\and
INAF - Osservatorio Astronomico di Roma, via di Frascati 33, Monte Porzio Catone, Italy\goodbreak
\and
INAF - Osservatorio Astronomico di Trieste, Via G.B. Tiepolo 11, Trieste, Italy\goodbreak
\and
INAF/IASF Bologna, Via Gobetti 101, Bologna, Italy\goodbreak
\and
INAF/IASF Milano, Via E. Bassini 15, Milano, Italy\goodbreak
\and
INFN, Sezione di Bologna, Via Irnerio 46, I-40126, Bologna, Italy\goodbreak
\and
INFN, Sezione di Roma 1, Universit\`{a} di Roma Sapienza, Piazzale Aldo Moro 2, 00185, Roma, Italy\goodbreak
\and
INFN, Sezione di Roma 2, Universit\`{a} di Roma Tor Vergata, Via della Ricerca Scientifica, 1, Roma, Italy\goodbreak
\and
INFN/National Institute for Nuclear Physics, Via Valerio 2, I-34127 Trieste, Italy\goodbreak
\and
IPAG: Institut de Plan\'{e}tologie et d'Astrophysique de Grenoble, Universit\'{e} Grenoble Alpes, IPAG, F-38000 Grenoble, France, CNRS, IPAG, F-38000 Grenoble, France\goodbreak
\and
Imperial College London, Astrophysics group, Blackett Laboratory, Prince Consort Road, London, SW7 2AZ, U.K.\goodbreak
\and
Infrared Processing and Analysis Center, California Institute of Technology, Pasadena, CA 91125, U.S.A.\goodbreak
\and
Institut Universitaire de France, 103, bd Saint-Michel, 75005, Paris, France\goodbreak
\and
Institut d'Astrophysique Spatiale, CNRS (UMR8617) Universit\'{e} Paris-Sud 11, B\^{a}timent 121, Orsay, France\goodbreak
\and
Institut d'Astrophysique de Paris, CNRS (UMR7095), 98 bis Boulevard Arago, F-75014, Paris, France\goodbreak
\and
Institute of Astronomy, University of Cambridge, Madingley Road, Cambridge CB3 0HA, U.K.\goodbreak
\and
Institute of Theoretical Astrophysics, University of Oslo, Blindern, Oslo, Norway\goodbreak
\and
Instituto de Astrof\'{\i}sica de Canarias, C/V\'{\i}a L\'{a}ctea s/n, La Laguna, Tenerife, Spain\goodbreak
\and
Instituto de F\'{\i}sica de Cantabria (CSIC-Universidad de Cantabria), Avda. de los Castros s/n, Santander, Spain\goodbreak
\and
Istituto Nazionale di Fisica Nucleare, Sezione di Padova, via Marzolo 8, I-35131 Padova, Italy\goodbreak
\and
Jet Propulsion Laboratory, California Institute of Technology, 4800 Oak Grove Drive, Pasadena, California, U.S.A.\goodbreak
\and
Jodrell Bank Centre for Astrophysics, Alan Turing Building, School of Physics and Astronomy, The University of Manchester, Oxford Road, Manchester, M13 9PL, U.K.\goodbreak
\and
Kavli Institute for Cosmology Cambridge, Madingley Road, Cambridge, CB3 0HA, U.K.\goodbreak
\and
LAL, Universit\'{e} Paris-Sud, CNRS/IN2P3, Orsay, France\goodbreak
\and
LERMA, CNRS, Observatoire de Paris, 61 Avenue de l'Observatoire, Paris, France\goodbreak
\and
Laboratoire AIM, IRFU/Service d'Astrophysique - CEA/DSM - CNRS - Universit\'{e} Paris Diderot, B\^{a}t. 709, CEA-Saclay, F-91191 Gif-sur-Yvette Cedex, France\goodbreak
\and
Laboratoire Traitement et Communication de l'Information, CNRS (UMR 5141) and T\'{e}l\'{e}com ParisTech, 46 rue Barrault F-75634 Paris Cedex 13, France\goodbreak
\and
Laboratoire de Physique Subatomique et de Cosmologie, Universit\'{e} Joseph Fourier Grenoble I, CNRS/IN2P3, Institut National Polytechnique de Grenoble, 53 rue des Martyrs, 38026 Grenoble cedex, France\goodbreak
\and
Laboratoire de Physique Th\'{e}orique, Universit\'{e} Paris-Sud 11 \& CNRS, B\^{a}timent 210, 91405 Orsay, France\goodbreak
\and
Lawrence Berkeley National Laboratory, Berkeley, California, U.S.A.\goodbreak
\and
Max-Planck-Institut f\"{u}r Astrophysik, Karl-Schwarzschild-Str. 1, 85741 Garching, Germany\goodbreak
\and
Max-Planck-Institut f\"{u}r Radioastronomie, Auf dem H\"{u}gel 69, 53121 Bonn, Germany\goodbreak
\and
McGill Physics, Ernest Rutherford Physics Building, McGill University, 3600 rue University, Montr\'{e}al, QC, H3A 2T8, Canada\goodbreak
\and
National University of Ireland, Department of Experimental Physics, Maynooth, Co. Kildare, Ireland\goodbreak
\and
Niels Bohr Institute, Blegdamsvej 17, Copenhagen, Denmark\goodbreak
\and
Observational Cosmology, Mail Stop 367-17, California Institute of Technology, Pasadena, CA, 91125, U.S.A.\goodbreak
\and
Optical Science Laboratory, University College London, Gower Street, London, U.K.\goodbreak
\and
P.N. Lebedev Physical Institute of the Russian Academy of Sciences, Astro Space Centre, 84/32 Profsoyuznaya st., Moscow, GSP-7, 117997, Russia\goodbreak
\and
SISSA, Astrophysics Sector, via Bonomea 265, 34136, Trieste, Italy\goodbreak
\and
School of Physics and Astronomy, Cardiff University, Queens Buildings, The Parade, Cardiff, CF24 3AA, U.K.\goodbreak
\and
School of Physics and Astronomy, University of Nottingham, Nottingham NG7 2RD, U.K.\goodbreak
\and
Space Sciences Laboratory, University of California, Berkeley, California, U.S.A.\goodbreak
\and
Special Astrophysical Observatory, Russian Academy of Sciences, Nizhnij Arkhyz, Zelenchukskiy region, Karachai-Cherkessian Republic, 369167, Russia\goodbreak
\and
Sub-Department of Astrophysics, University of Oxford, Keble Road, Oxford OX1 3RH, U.K.\goodbreak
\and
UPMC Univ Paris 06, UMR7095, 98 bis Boulevard Arago, F-75014, Paris, France\goodbreak
\and
Universit\'{e} de Toulouse, UPS-OMP, IRAP, F-31028 Toulouse cedex 4, France\goodbreak
% ...
}

\title{\Planck\ intermediate results. XXXIV. The magnetic field
  structure in the Rosette Nebula}
\date{Received 05 January 2015/ Accepted 08 April 2015}

  \abstract
  { \Planck\ has mapped the polarized dust emission over the whole sky,
    making it possible to trace the Galactic magnetic field structure that
    pervades the interstellar medium (ISM). We combine polarization
    data from \Planck\ with rotation 
    measure (RM) observations towards a massive star-forming region, the Rosette
    Nebula in the Monoceros molecular cloud, to study its
    magnetic field structure and the impact of an expanding
    \hii~region on the morphology of the field. We derive an
    analytical solution for the magnetic field, assumed to
    evolve from an initially uniform configuration following the
    expansion of ionized gas and the formation of a shell of swept-up
    ISM. From the RM data we estimate a mean value of the
    line-of-sight component of the magnetic field of about 
    3\,$\mu$G (towards the observer) in the Rosette Nebula, for a
    uniform electron density of about 12\,cm$^{-3}$.
    The dust shell that surrounds the Rosette \hii~region
    is clearly observed in the \Planck\ intensity map at
    353\,GHz, with a polarization signal significantly different from
    that of the local background when considered as a whole. The
    \Planck\ observations constrain the plane-of-the-sky orientation
    of the magnetic field in the Rosette's parent molecular cloud to
    be mostly aligned with the large-scale field along the 
    Galactic plane. The \Planck\ data are compared with the
    analytical model, which predicts the mean polarization properties of a
    spherical and uniform dust shell for a given orientation of the
    field. This comparison leads to an upper limit of about 45\degr\ on the angle
    between the line of sight and the magnetic field in the Rosette
    complex, for an assumed intrinsic dust polarization fraction of 4\,\%.
    This field direction can reproduce the RM values detected in the
    ionized region if the magnetic field strength in the Monoceros
    molecular cloud is in the range 6.5--9\,$\mu$G. The present
  analytical model is able to reproduce the
    RM distribution across the ionized nebula, as well as the mean dust
    polarization properties of the swept-up shell, and can be directly
    applied to other similar objects. }

\keywords{Polarization -- ISM: magnetic fields -- radiation mechanisms: general -- radio continuum: ISM -- submillimeter: ISM}
 
\titlerunning{The magnetic field structure in the Rosette Nebula}
\authorrunning{Planck Collaboration}
\maketitle

\section{Introduction}
\label{sec:intro}

Stellar bubbles are formed from the combined action of ionization power
and stellar winds from OB stars. They are a major source of turbulent
energy injection into the interstellar medium (ISM), sweeping up the surrounding gas and
dust and modifying the magnetic fields (e.g. \citealt{Ferriere:2001}). Bubbles are
seen as agents of triggered star formation, as the
swept-up material may become unstable and fragment. It is therefore
important to understand the influence of star formation and feedback
on the structure of the magnetic field, and how the ionized
bubbles are shaped by the field. The all-sky polarization data from
\Planck\footnote{\Planck\ (\url{http://www.esa.int/Planck}) is a
  project of the European Space Agency (ESA) with instruments
  provided by two scientific consortia funded by ESA member states (in
  particular the lead countries France and Italy), with contributions
  from NASA (USA) and telescope reflectors provided by a collaboration
  between ESA and a scientific consortium led and funded by Denmark.}
open up new opportunities for studying the magnetic field structure
of such objects and their connection with the large-scale Galactic
field. 

The interplay between the
expansion of an ionized nebula and the action of the
magnetic fields has been the subject of several studies, both
observational (e.g. \citealt{Pavel:2012}, \citealt{Santos:2012},
\citealt{Santos:2014}) and numerical (e.g. \citealt{Bernstein:1965},
\citealt{Giuliani:1982}, \citealt{Ferriere:1991},
\citealt{Krumholz:2007},
\citealt{Peters:2011},\citealt{Arthur:2011}). These observational
studies rely partly on interstellar 
dust grains as tracers of the magnetic field. Non-spherical grains spin around
their axes of maximal inertia, which are precessing around the magnetic field
lines. The grains emit preferentially along their long axes, thus
giving rise to an electric vector perpendicular to the magnetic field.
At the same time, dust grains polarize the light from background
stars and since extinction is higher along their longest axes, the
transmitted electric vector is parallel to the magnetic field in the
plane of the sky (see e.g. \citealt{Martin:2007}).
Observations of dust emission or absorption allow
us to retrieve the plane-of-the-sky orientation of the magnetic field that pervades interstellar
matter. Results from high-resolution studies of \hii~regions have
revealed regions of well-ordered magnetic field along the edges of
the nebulae and magnetic field strengths of tens to hundreds of
$\mu$G. \citet{Santos:2014} studied the star-forming region
Sh2-29 in optical and near-infrared polarimetry and derived a field
strength of around $400$\,$\mu$G. The authors used the Chandrasekhar-Fermi method
\citep{Chandrasekhar:1953}, which relates the dispersion in magnetic
field orientation with turbulent motions of the gas, under the
assumption of equipartition between magnetic and turbulent/thermal pressures. Such high values
of the field strength reflect the ordered structure of the magnetic
field at sub-parsec scales, compressed by the expanding
\hii~region. The aforementioned dust polarization observations of
ionized nebulae, both in emission and
extinction, have covered small regions around the
objects, generally tens of arcminutes. \Planck\ data allow us to study
the magnetic fields probed by polarized dust emission towards large
\hii~regions embedded in their parent molecular clouds, and the
diffuse medium surrounding them.

Measurements of the magnetic fields towards \hii~regions can also be
performed through Faraday rotation of linearly polarized background sources, both
extragalactic and Galactic \citep{Heiles:1980}. The plasma in the \hii~regions rotates the plane
of polarization of the background radio wave by an angle that is
proportional to the square of the observing wavelength. The quantity
derived from Faraday rotation observations, the rotation
measure (RM), is directly related to the line-of-sight component of
the magnetic field $B_{||}$ weighted by the electron density of the ionized
gas. \citet{Harvey-Smith:2011} studied the line-of-sight magnetic field towards five large diameter Galactic
\hii~regions, tens of parsecs wide, using RM combined with \ha\ data
to estimate the electron density. They found $B_{||}$ values of 2--6\,$\mu$G,
consistent with those measured in the diffuse ISM through Zeeman
splitting observations \citep{Heiles:2005,Crutcher:2010,Crutcher:2012}. This indicates that the RM
enhancement observed towards \hii~regions may be the consequence of a local increase of
the electron density. A similar study performed on the Rosette nebula
by \citet{Savage:2013} attributed the high values of RM to an increase
in the field strength, as will be
discussed in Sect. \ref{sec:intro2}. 

The goal of this work is to study the structure of the magnetic
field in and around an ionized bubble created by young stars. For this
purpose we chose
the Rosette nebula, adjacent to the
Monoceros (Mon) OB2
cloud, for its close to spherical shape and for its relatively large
size on the sky, $1\fdg4$, relative to the \Planck\ 353\,GHz
beam of $4\parcm9$. This paper is
organized as follows. We start in Sect. \ref{sec:intro2} by describing the Rosette nebula and
its main features relevant for our study. In
Sect. \ref{sec:data} we introduce the \Planck\ and ancillary data used
in this study, which are then analysed and discussed in
Sect. \ref{sec:obs}. The interpretation of the radio and
submillimetre polarization
observations is presented in Sect. \ref{sec:magfield}, in light of a 2D 
analytical model of the magnetic field in a spherical bubble-shell
structure. The
main results are summarized in Sect. \ref{sec:conc}. The detailed 
derivation of our magnetic field model is given in Appendix \ref{appb}.
\begin{figure}
\centering
\includegraphics[scale=0.45]{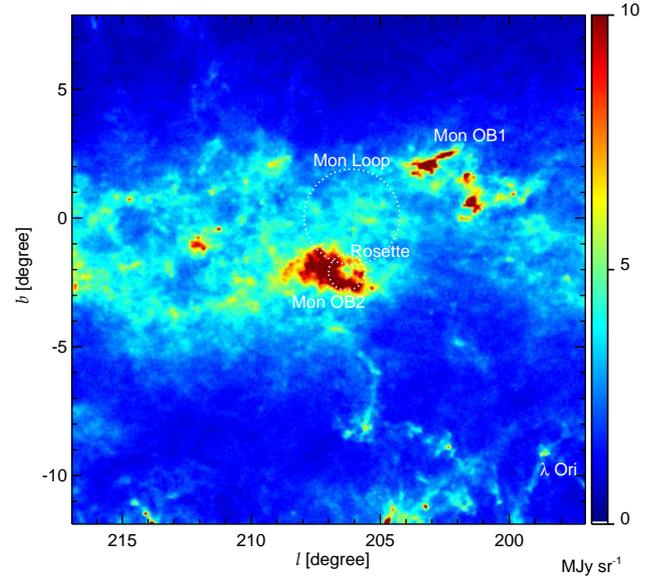}
\caption{The Rosette Nebula, its molecular cloud Mon OB2, and their
  surrounding medium as seen by \Planck\ at 353\,GHz. The dotted
  lines give the approximate outer radii of the Rosette \hii~region
  and the Mon Loop supernova remnant (described in
  Sect. \ref{sec:intro2}). The northern Monoceros cloud Mon 
  OB1 is above the Galactic plane; part of the dust ring of $\lambda$
Ori is visible at the bottom right corner of the map; the northern end
of Orion B is seen at $(l,b)=(204\degr,-12\degr)$ \citep{Dame:2001}.}
\label{fig:largemap}
\end{figure}

\section{The Rosette Nebula and the Mon OB2 molecular cloud}
\label{sec:intro2}

The Rosette bubble is located near the anti-centre of the Galactic
disk, centred on $(l,b)=(206\pdeg3,-2\pdeg1)$, and is part of a larger region called
the northern Monoceros region. It is at
($1.6\pm0.1$)\,kpc from the Sun and its age is estimated to be about $(3\pm1)$\,Myr
\citep{Roman:2008}. Figure \ref{fig:largemap} shows the position of
the Rosette Nebula relative to its parent molecular cloud Mon OB2, as
well as other known and nearby objects, marked on the \Planck\ 353\,GHz map
(Sect. \ref{sec:planckd}). The Rosette has been the object of
extensive studies, as it has been considered an archetype of a
triggered star formation site
\citep{Williams:1994,Balog:2007,Roman:2008,Schneider:2010}. However,
based on the small age spread among the young stellar objects in the
Rosette, \citet{Ybarra:2013} and \citet{Cambresy:2013} suggest that the
effect of the \hii~region expansion on stimulating star formation is secondary
relative to the original collapse of the cloud. The prominent OB
association, NGC 2244, responsible for the ionized nebula and the
evacuation of its central part, contains more than 100 stars, of which
seven are O and
24 are B stars \citep{Ogura:1981,Marschall:1982,Park:2002,Roman:2008}. \citet{Cox:1990}
quote a luminosity of
$22\times10^{5}$\,L$_{\odot}$ for the brightest four O stars and 13 B
stars of the central cluster.

The first study of the expansion of the \hii~region in the Rosette was
done by \citet{Minkowski:1949}. The author also noted the existence of
``elephant trunks'' and dark globules, from the analysis of photographic
data, in the north-western edge of the central
cavity \citep{Herbig:1974, Schneps:1980,Carlqvist:1998,
  Carlqvist:2002}. Globules are also seen in the south-eastern region
in the \textit{Herschel} images of the Rosette
\citep{Schneider:2010}, where the ionized nebula
is interacting with the molecular cloud. The Rosette \hii~region is an ionization bounded
Str\"{o}mgren sphere \citep{Menon:1962} and expands at a velocity of
$\lsim 15$\kms\ \citep{Smith:1973,Fountain:1979,Meaburn:1968a}. 
The mean electron temperature in the nebula, derived by \citet{Quireza:2006}, is 8500\,K.

The Rosette \hii~region has a projected optical extent
of about $1\fdg4$. 
\citet{Celnik:1985} estimated the size of the nebula by fitting
several shell models to observations at 1.4 and 4.7\,GHz. The mean
radial profile of the radio emission was reproduced by a model of a spherical
shell with inner and outer radii of 7 and 19\,pc, respectively, and
constant electron density of 13.5\,cm$^{-3}$ (these values have been scaled
from the distance of 1.42\,kpc used by \citeauthor{Celnik:1985} to the
adopted distance of 1.6\,kpc). \citeauthor{Celnik:1985} also found
that the ionized shell shows a radial density gradient out to an
outer radius of 28\,pc, with a mean electron density of about
5.7\,cm$^{-3}$. This model leads to a total mass of ionized
matter of $1.2\times10^{4}$\,M$_{\odot}$, similar to the value of $1.3\times10^{4}$\,M$_{\odot}$
obtained for the spherical shell of constant density. 
The total molecular mass
derived from $^{12}$CO emission is $1.6\times10^{5}$\,M$_{\odot}$
\citep{Heyer:2006}, of which 54\,\% lies inside the Rosette
\hii~region. 
 
The magnetic field in the Rosette nebula has been recently studied by
\citet{Savage:2013} using RM observations of background galaxies. They
fit the RM data with a simple stellar bubble model
\citep{Whiting:2009} that consists of an inner low-density
cavity of shocked stellar wind and a shell of shocked and photoionized
ISM material. 
The authors assume that the ambient magnetic field \vec{B}$_{0}$ outside
the bubble has a strength $B_{0}=4$\,$\mu$G and
that the field component in the shock plane is amplified by a factor
of 4, appropriate for strong adiabatic shocks. Then by fitting the RM
data to their model, they find that the angle between \vec{B}$_{0}$ and
the line of sight is $\theta_{0}=$72\degr. The results of
\citet{Savage:2013} will be further discussed in Sect. \ref{sec:modelrm}.

Part of the northern Monoceros region has also been studied in
polarization at radio frequencies by \citet{Xiao:2012}. These authors
focused on the supernova remnant (SNR) Monoceros Loop, G205.7+0.1,
north-west of the Rosette nebula. The SNR and \hii~region are thought to be
interacting
\citep{Davies:1978,Smith:1973,Jaffe:1997,Fountain:1979,Gosachinskii:1982}. The
polarization data presented by \citet{Xiao:2012} do not
cover the interface region between the two objects and show
that the magnetic field in the Rosette Nebula is largely
  parallel to the Galactic plane. However, at a frequency of 5\,GHz
the observed polarization vectors are 
the result of either highly rotated background polarized synchrotron emission,
considering the high RMs of about 500\,rad\,m$^{-2}$ measured in the
nebula (Sect. \ref{sec:polrm}, \citealt{Savage:2013}), or foreground emission. 

\section{Data}
\label{sec:data}

\subsection{\Planck\ data}
\label{sec:planckd}

\Planck\ \citep{tauber2010a, planck2011-1.1,planck2013-p01} has mapped
the polarization of the sky emission in seven frequency channels, from
30 to 353\,GHz. In this paper we use the data from the High Frequency
Instrument (HFI; \citealt{Lamarre2010, planck2011-1.5,
  planck2013-p03}) at 353\,GHz, where the dust polarized emission is
brightest. A first description of the dust polarization sky is
presented in a series of papers, namely 
\citet{planck2014-XIX}, \citet{planck2014-XX}, \citet{planck2014-XXI},
\citet{planck2014-XXX}, \citet{planck2014-XXXII}, and \citet{planck2014-XXXIII}. 

\subsubsection{Intensity and polarization}
\label{sec:planckd1}

We use the full-mission temperature and polarization sky maps at
353\,GHz from the 2015 release of
\Planck\ \citep{planck2014-a01}. The map-making, calibration and
correction of systematic effects is
described in \citet{planck2014-a09}. 
We smooth the maps and their
corresponding covariance from the
initial angular resolution of $4\farcm9$ to 6\arcm. 
This is a compromise between increasing the signal-to-noise ratio of
the polarization data and preserving high resolution, as well as minimizing
beam depolarization effects.

The 353\,GHz intensity data are corrected for the cosmic microwave
background (CMB) anisotropies
using the {\tt Commander} map, presented in \citet{planck2014-a11}, although the contribution to the
total signal in the region under study is small, at most 2\,\%. We
also subtract the Galactic zero level offset,
$0.0885\pm0.0067$\,MJy\,sr$^{-1}$, from the intensity map and add the
corresponding uncertainty to the intensity variance
\citep{planck2014-a09}. 
We do not attempt to correct for the
polarized signal of the CMB anisotropies, as this is a negligible
contribution $\ll 1$\,\% at 353\,GHz in the region under study \citep{planck2014-XXX}.

The linear polarization of dust emission is measured from the Stokes
parameters $Q$, $U$, and $I$, delivered in the \Planck\ data release. They are
the result of a line-of-sight integration and are related as
\begin{align}
\label{poleqs}
& Q  = I p  \cos(2\psi), \nonumber \\
& U  = I p  \sin(2\psi),   \nonumber \\
& p = \frac{P}{I} = \frac{\sqrt{Q^{2} + U^{2}}} {I},  \nonumber \\
& \psi = 0.5 \arctan(U, Q),
\end{align}
where $p$ is the dust polarization fraction 
and $\psi$ is the polarization angle.
The two argument function $\arctan(X,Y)$ is used to compute
${\arctan}(X/Y)$, avoiding the $\pi$ ambiguity. The \Planck\ Stokes
parameters are provided in the {\tt HEALPix} \citep{Gorski:2005}
convention, such that the angle $\psi=0\degr$ is towards the north Galactic pole and
positive towards the west (clockwise). In the commonly used IAU
convention \citep{Hamaker:1996}, the polarization angle is measured
anticlockwise relative to the north Galactic pole. We adopt the IAU
convention by taking the negative of the \Planck\ $U$ Stokes map. 

The Stokes $Q$ and $U$ maps in the region under study have a
  typical signal-to-noise ratio of 8 and 2, respectively. The main systematic effect
concerning polarization data is signal leakage from total
intensity. We used the corrections derived from the global template fit,
described in \citet{planck2014-a09}, which accounts for all of the
leakage sources, namely bandpass, monopole, and calibration
mismatches. The 2015 \Planck\ data release also includes polarization
maps that were corrected only for dust bandpass leakage using
a different method. We compare the two sets of maps in the region under
study to quantify the systematic uncertainties associated with the
leakage correction. The relative difference in the $Q$ Stokes parameter is less
than 10\,\%, whereas it is typically 25\,\% in $U$ because the Rosette is 2\degr\ from the Galactic plane, where
the magnetic fields are largely parallel to the plane and contribute
more signal to $Q$ than to $U$.

Because of the presence of noise in the data, the polarization
  intensity, $P$, calculated directly using Eq. (\ref{poleqs}) is
  positively biased. We debias this quantity according to the method proposed
by \citet{Plaszczynski:2014} by taking into account the diagonal and
off-diagonal terms in the covariance matrix for $Q$ and $U$. Since we
only calculate the polarization
  intensity towards high ($>10$) signal-to-noise regions, namely the
  Mon OB2
  cloud, the relative difference between the debiased and direct
  (Eq. \ref{poleqs}) estimates
  is less than 3\,\%. A comparison between other debiasing methods as applied to the
all-sky \Planck\ data is presented in \citet{planck2014-XIX}.

\subsubsection{Products}
\label{sec:planckd2}

We use the \Planck\  ``Type 3" CO map from \citet{planck2013-p03a}
to qualitatively inspect the molecular gas in the Monoceros region. This is a
composite line map, where the line ratios between the first three CO
rotational lines are assumed to be constant across the whole sky. The
Type 3 CO map is the highest signal-to-noise map extracted from
\Planck, with an angular resolution of $5\parcm5$. 

The dust emission from \Planck\ wavelengths to about 100\,$\mu$m is dominated
  by the contribution of large dust grains (radius
  larger than 0.05\,$\mu$m). In order to trace their
  temperature, we use the map derived from the all-sky model of dust emission
from \citet{planck2013-p06b}, at a resolution of $5\arcm$. 

\subsection{RM data}
\label{sec:rmd}

We use the RM data presented by \citet{Savage:2013} towards 23 extragalactic
radio sources observed through the Rosette complex with the Karl
G. Jansky Very Large Array. \footnote{We checked for the
    presence of nearby pulsars in the Australia Telescope National Facility Pulsar
  Catalogue \citep{Manchester:2005}. We did not find pulsars closer
  than $2\degr$ from the centre of the Rosette with the required information (distance,
  dispersion measure, RM).}
The RM observations towards extragalactic sources are an integral of
the line-of-sight magnetic field component $B_{||}$, weighted by the electron
density $n_{\rm e}$, and given by
\begin{equation}
\label{rmeq}
{\rm RM}= 0.81 \int_{0}^{S} \left(\frac{n_{\rm e}}{\rm
    cm^{-3}}\right) \left(\frac{B_{||}}{\mu \rm G}\right)
\left(\frac{ds}{\rm pc}\right) {\rm rad\,m^{-2}},
\end{equation}
where $S$ is the path length from the source to the observer. The RM is positive when
\vec{B} points towards us, hence when $B_{||}$ is also
positive. 

We consider 20
measurements (given in Table 3 of \citealt{Savage:2013}); two of the sources are depolarized
and we also exclude the only negative RM value detected. This
measurement was also discarded in the analysis of \citet{Savage:2013}, who
could not determine if its origin is Galactic or extragalactic. In
any case this measurement lies outside the boundaries of
the Rosette Nebula. 
Whenever two values are given for the same
source, in the case of a double component extragalactic object, we
take their mean and the dispersion as the
uncertainty because we combine the RM data at a
resolution of 12\parcs8, with emission measure data at a lower
resolution of 14\farcm4 (Sect. \ref{sec:emd}). The 20 RM
  measurements are positive and have a typical uncertainty
of 40\,\%, or below 10\,\% for half of the sample.

\subsection{Emission measure data}
\label{sec:emd}

The emission measure (EM) data are from the radio recombination line
(RRL) survey of \citet{Alves:2014} at 1.4\,GHz and at a resolution of
$14\farcm4$. These are a by-product of the \hi\ Parkes All-Sky Survey
(HIPASS, \citealt{Staveley-Smith:1996}) and their analysis is
presented by \citet{Alves:2010, Alves:2012}. 
The emission measure is defined as 
\begin{equation}
\label{emeq1}
{\rm EM} = \int_{0}^{\infty} n_{\rm e}^{2} ds
\end{equation}
and derived from the integrated RRL emission as
\begin{equation}
\label{emeq2}
{\rm EM} = 5.2\times 10^{-4} \left(\frac{T_{\rm e}}{\rm K}\right)^{1.5} \int \left(\frac{T_{\rm
  L} d\nu}{\rm K\,kHz}\right) {\rm cm^{-6} \rm pc},
\end{equation}
where $T_{\rm e}$ and $T_{\rm L}$ are the electron and line
temperatures, respectively. The overall calibration uncertainty in
these data is 10\,\%.

\subsection{IRAS data}
\label{sec:irasd}

We use the IRIS (Improved Reprocessing of the IRAS Survey, \citealt{Miville-Deschenes:2005})
data at 12 and 100\,$\mu$m. The IRAS map at 12\,$\mu$m traces
  the emission from dust particles that are smaller than those
emitting at the longer \Planck\ wavelengths. 

\section{Our observational perspective on the Rosette}
\label{sec:obs}

In this section we start by describing the features of the Rosette Nebula and its parent
molecular cloud from total intensity maps at different frequencies. We then study
the polarized emission arising from the dust shell that surrounds the Rosette \hii~region, 
along with the radio polarization signal created by the ionized gas.

\begin{figure*}
\centering
\includegraphics[scale=0.4]{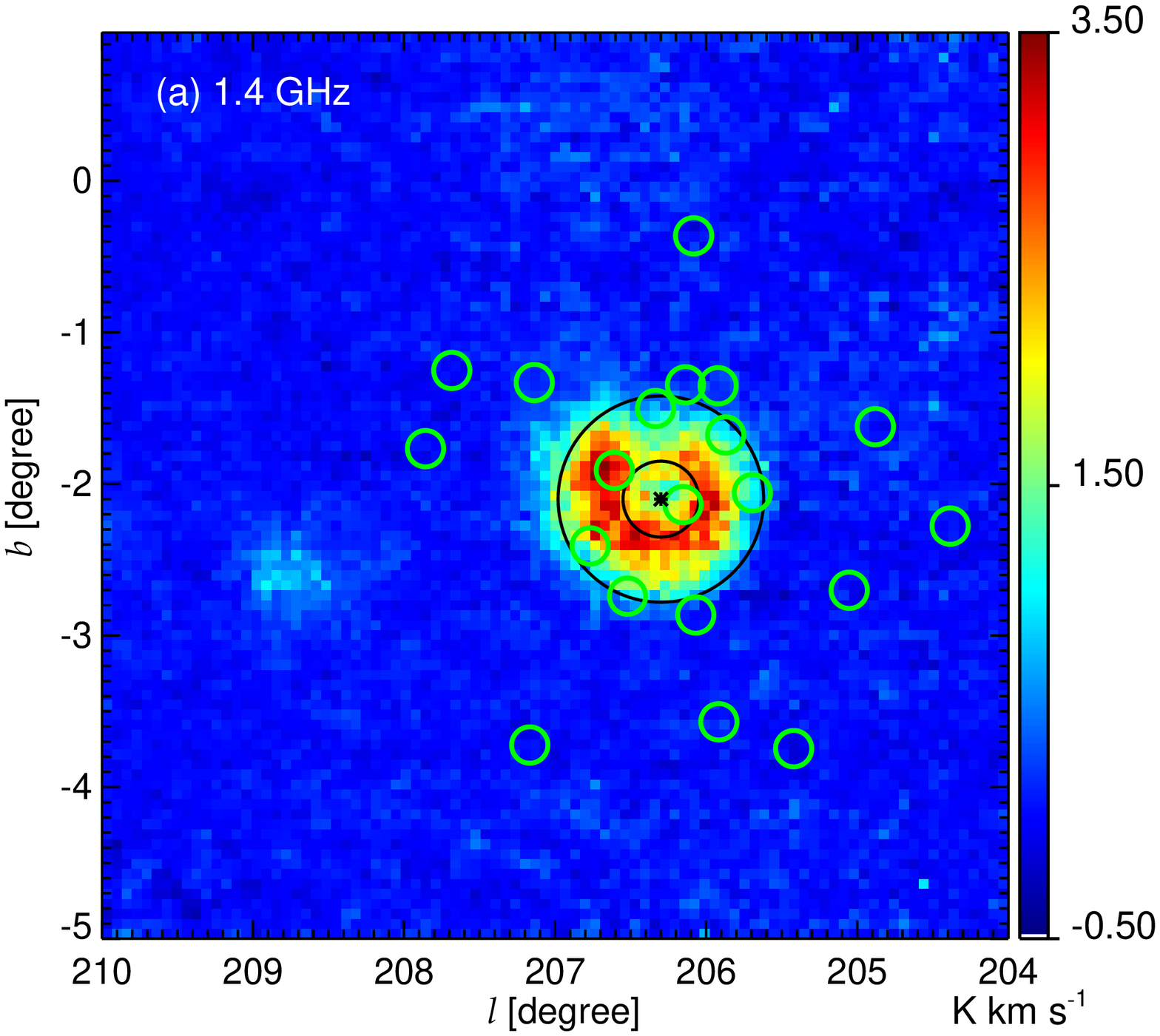}\vspace{-0.2cm}\hspace{0.2cm}
\includegraphics[scale=0.4]{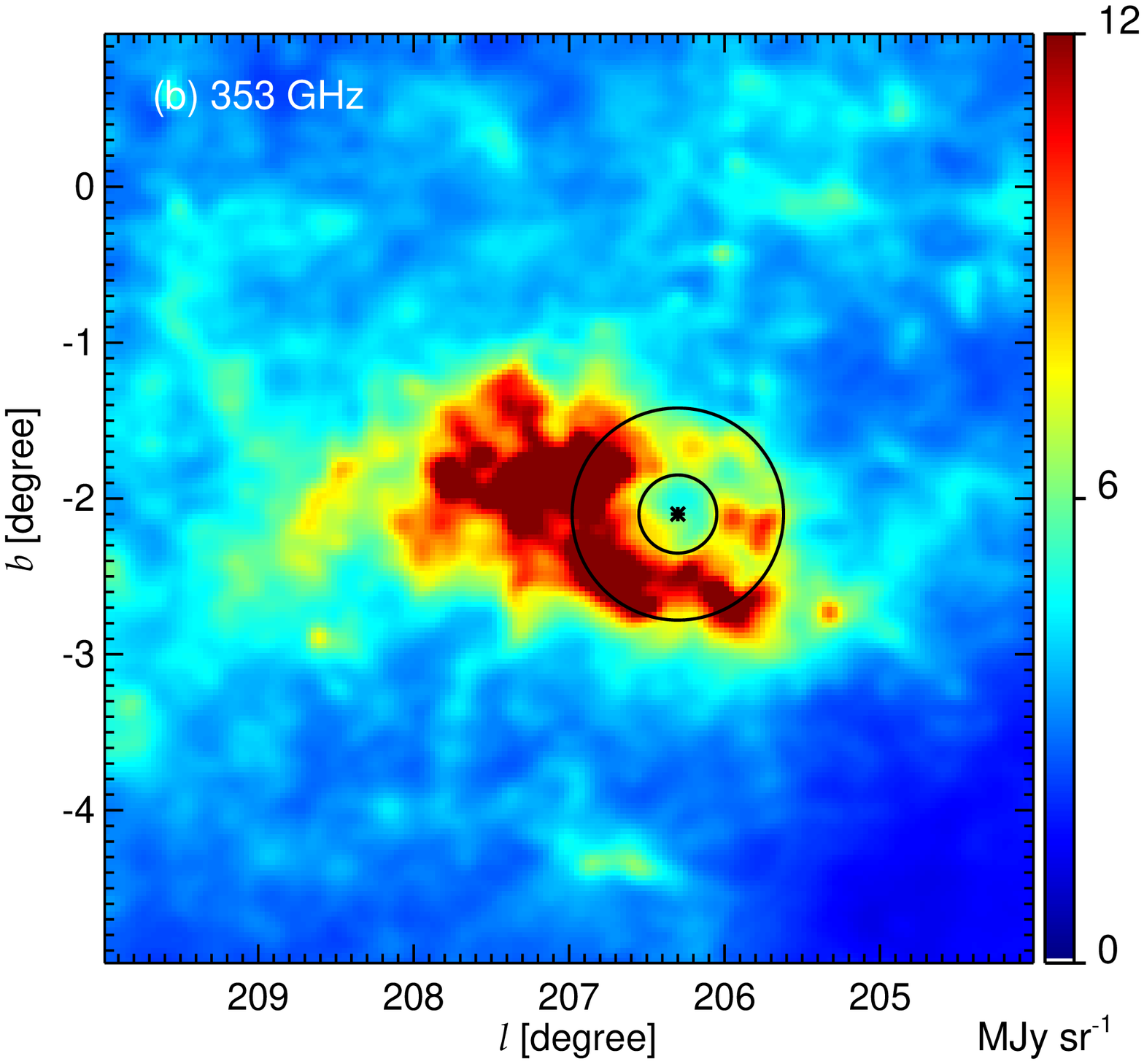}\vspace{-0.2cm}
\includegraphics[scale=0.4]{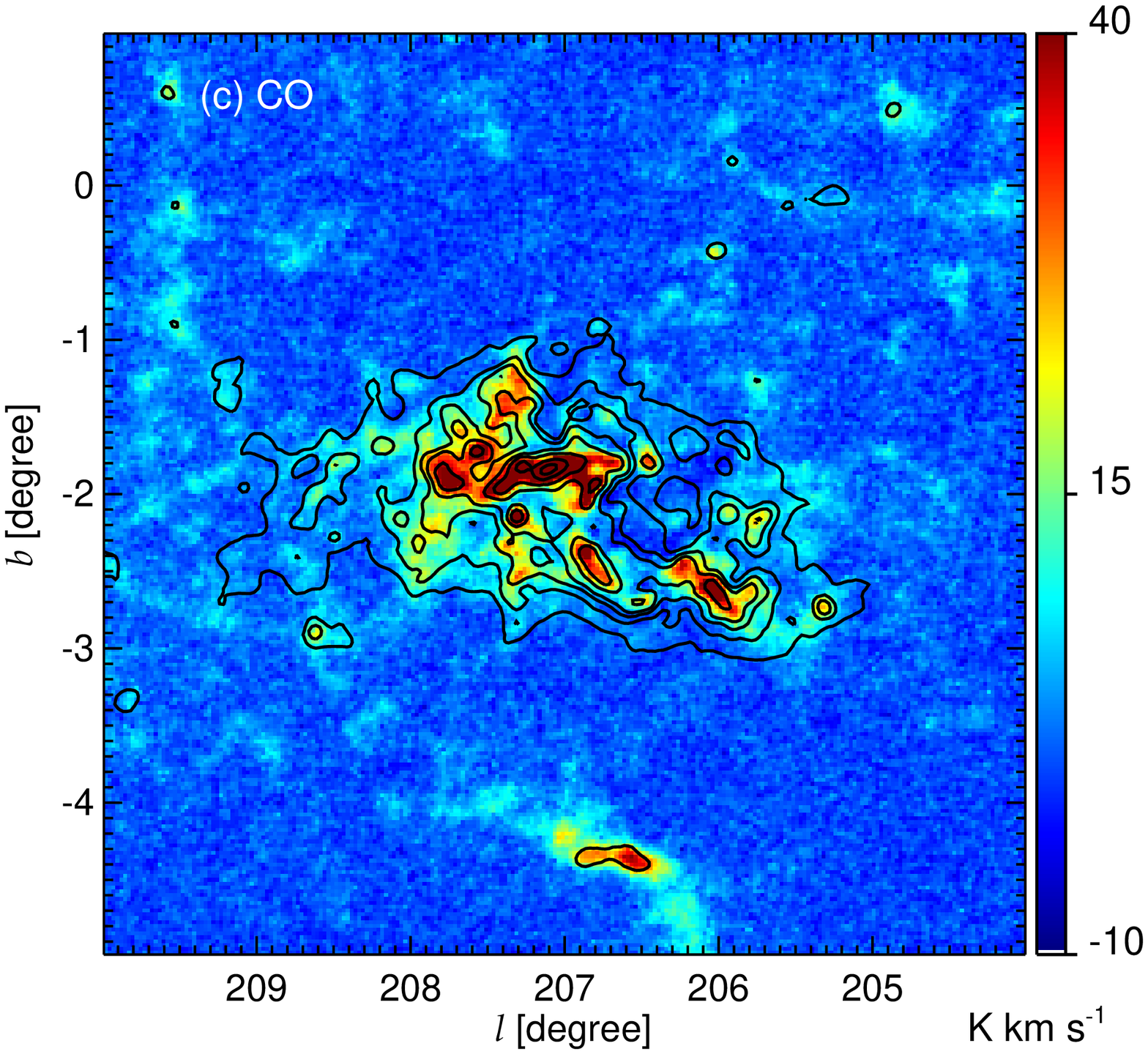}\vspace{-0.2cm}\hspace{0.2cm}
\includegraphics[scale=0.4]{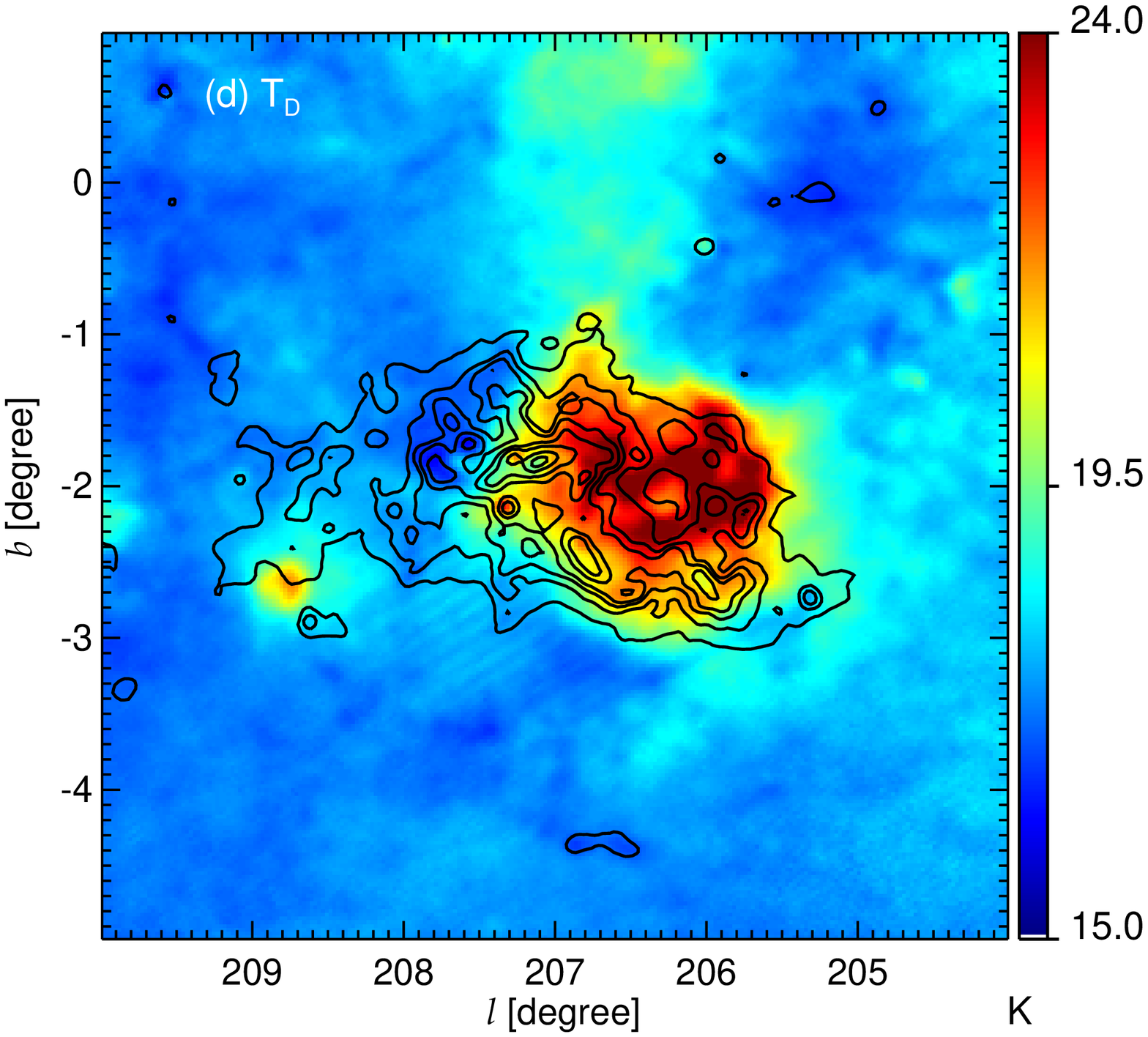}\vspace{-0.2cm}
\includegraphics[scale=0.4]{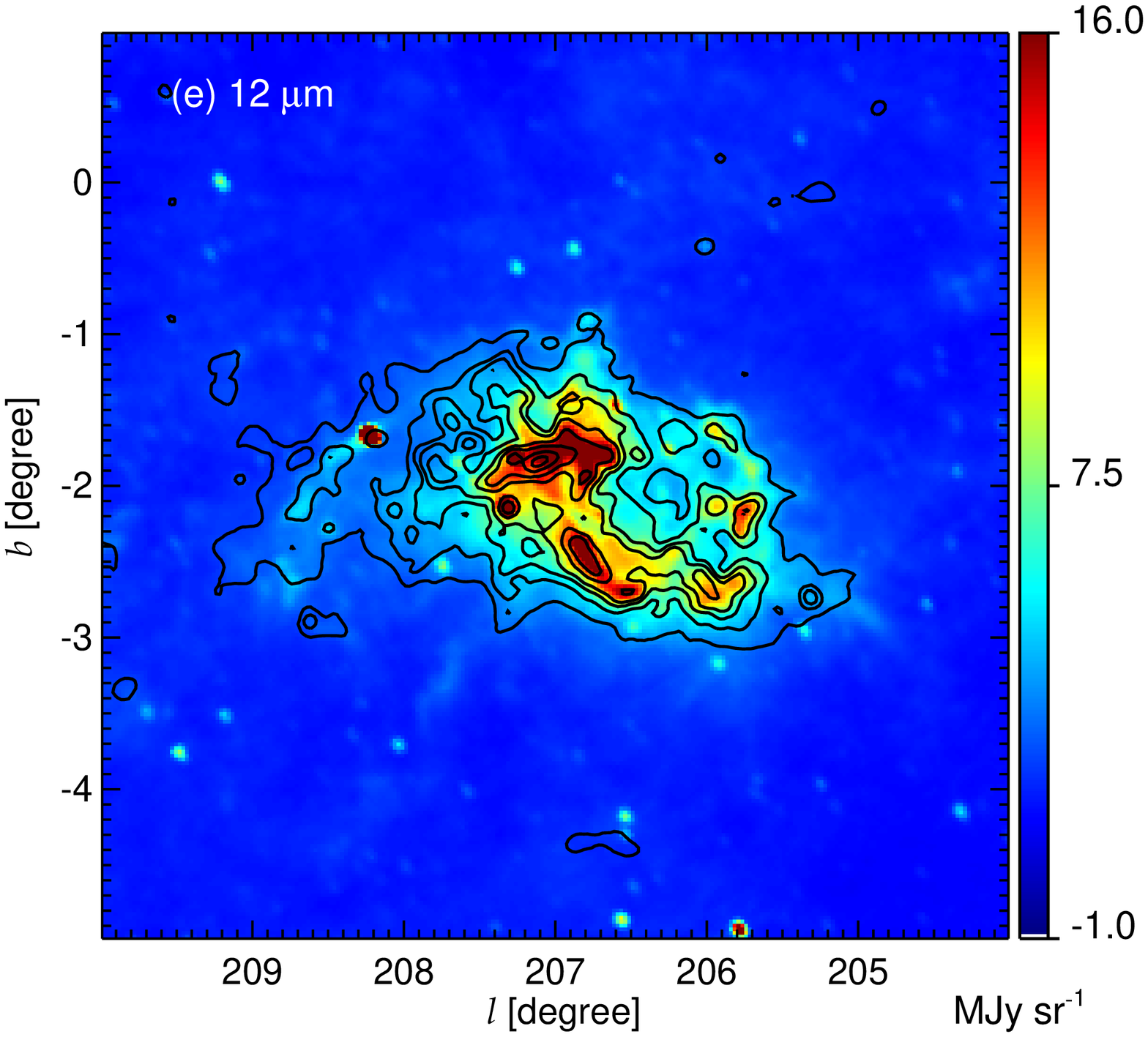}\vspace{-0.2cm}\hspace{0.2cm}
\includegraphics[scale=0.4]{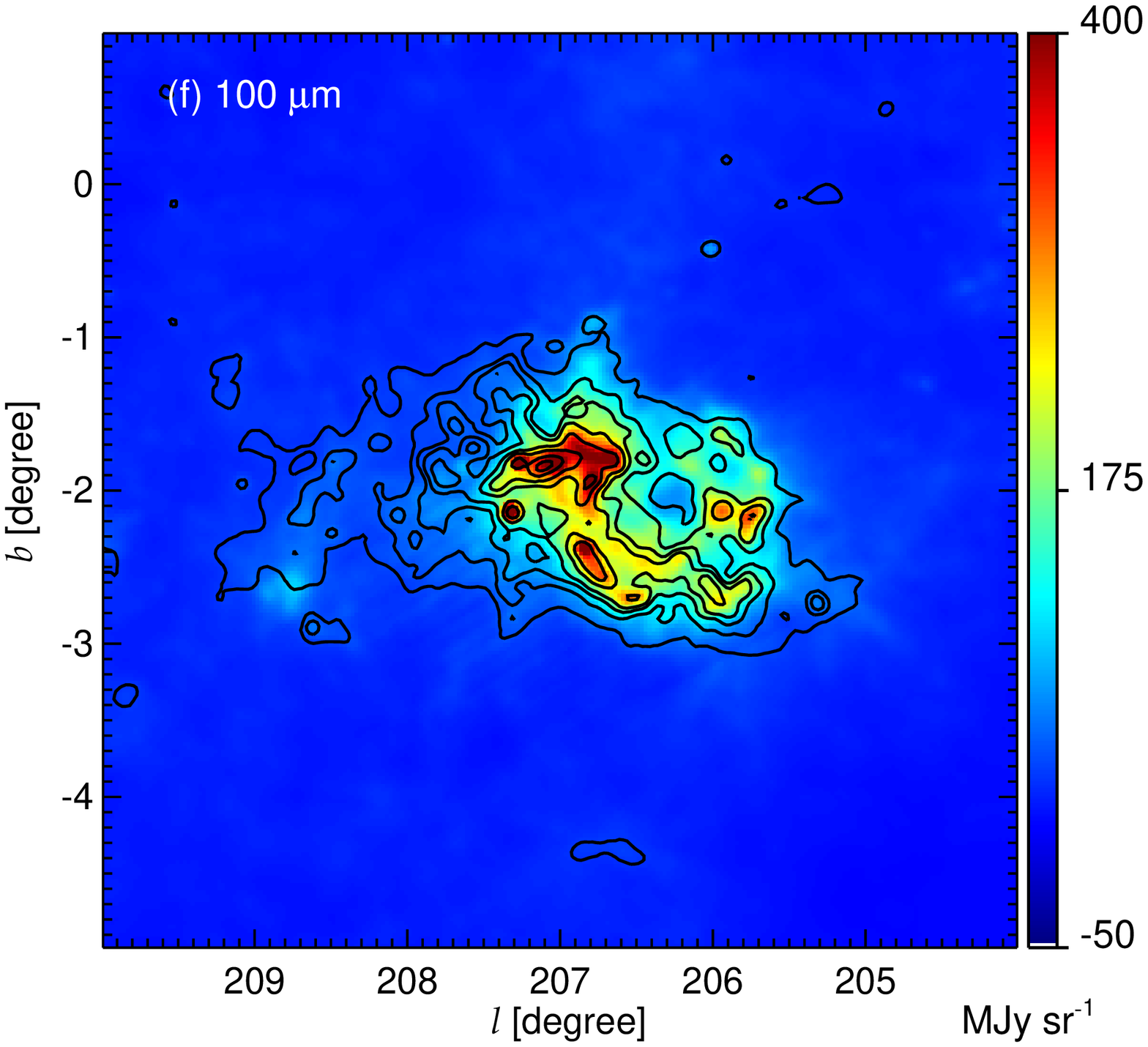}
\caption{The Rosette Nebula and its molecular cloud. (a) Integrated
  RRL emission at $14\farcm4$ resolution, showing the radio morphology
  of the \hii~region. The green circles,
  with a diameter equivalent to the beam full width
at half maximum (FWHM) of the RRL data, give the
  position of the 20 RM observations of \citet{Savage:2013} used in
  this work. (b)
  \Planck\ 353\,GHz emission at 6\arcm\ resolution. The black star in panels (a) and
  (b) indicates the position of the central star cluster NGC 2244 and the black circles correspond to the
  inner and outer radii of the \hii~region, 
 $r_{\rm in}^{\hii}=7$\,pc and $r_{\rm
    out}^{\hii}=19$\,pc, respectively (see Sect. \ref{sec:polrm}). (c) \Planck\ CO
  integrated intensity at $5\farcm5$ resolution. (d) Dust temperature derived
  from \Planck\ at 5\arcm\ resolution. (e) Dust emission
as seen by IRAS at 12\,$\mu$m and at 100\,$\mu$m (f), at
4\arcm\ resolution. The contours in panels (c) to (f) show
the \Planck\ 353\,GHz emission; these are at every 5\,\% from 15 to
30\,\%, at every 10\,\% from 30 to
50\,\%, and finally at 70 and 90\,\% of the maximum intensity of 37\,MJy\,sr$^{-1}$.}
\label{fig:map1}
\end{figure*}

\subsection{Intensity}
\label{sec:int}

The Rosette Nebula and its parent molecular cloud are presented in
Fig. \ref{fig:map1}. The maps show the RRL thermal radio emission, dust emission at
353\,GHz, integrated CO emission, dust temperature, and
dust emission at 12\,$\mu$m and 100\,$\mu$m. The central cavity is seen both in radio and in dust
emission, as a lower intensity region around
the position of the ionizing source NGC 2244. 

The expanding
\hii~region is interacting with the molecular cloud Mon OB2, creating
photon-dominated regions at their interface 
\citep{Cox:1990,Schneider:2010}, which can be traced at
12\,$\mu$m. The map of 
Fig. \ref{fig:map1}(e) shows bright 12\,$\mu$m emission associated
with the 353\,GHz contours on the eastern side of the
nebula's centre. Within these structures elongated
along the ionization front, there are dense molecular clumps with
on-going star formation, as the
result of the compression by the \hii~region. The dust temperature is
also higher at the interface of these regions with the nebula, as
shown in Fig. \ref{fig:map1}(d), compared to the lower temperature at
the position of their maximum intensity.

\citet{Cox:1990} studied the Rosette complex in all four IRAS
bands and analysed the radial distribution of the dust emission. These
authors showed that the 12\,$\mu$m emission, which traces the
photodissociation regions, comes from a shell surrounding the
outer side of the ionization front, as well as from the molecular
cloud. On the other hand, the longer wavelength emission at 60 and
100\,$\mu$m mostly arises from the \hii~region and thus is
tracing ionized and neutral gas.
\begin{figure*}
\centering
\includegraphics[scale=0.4]{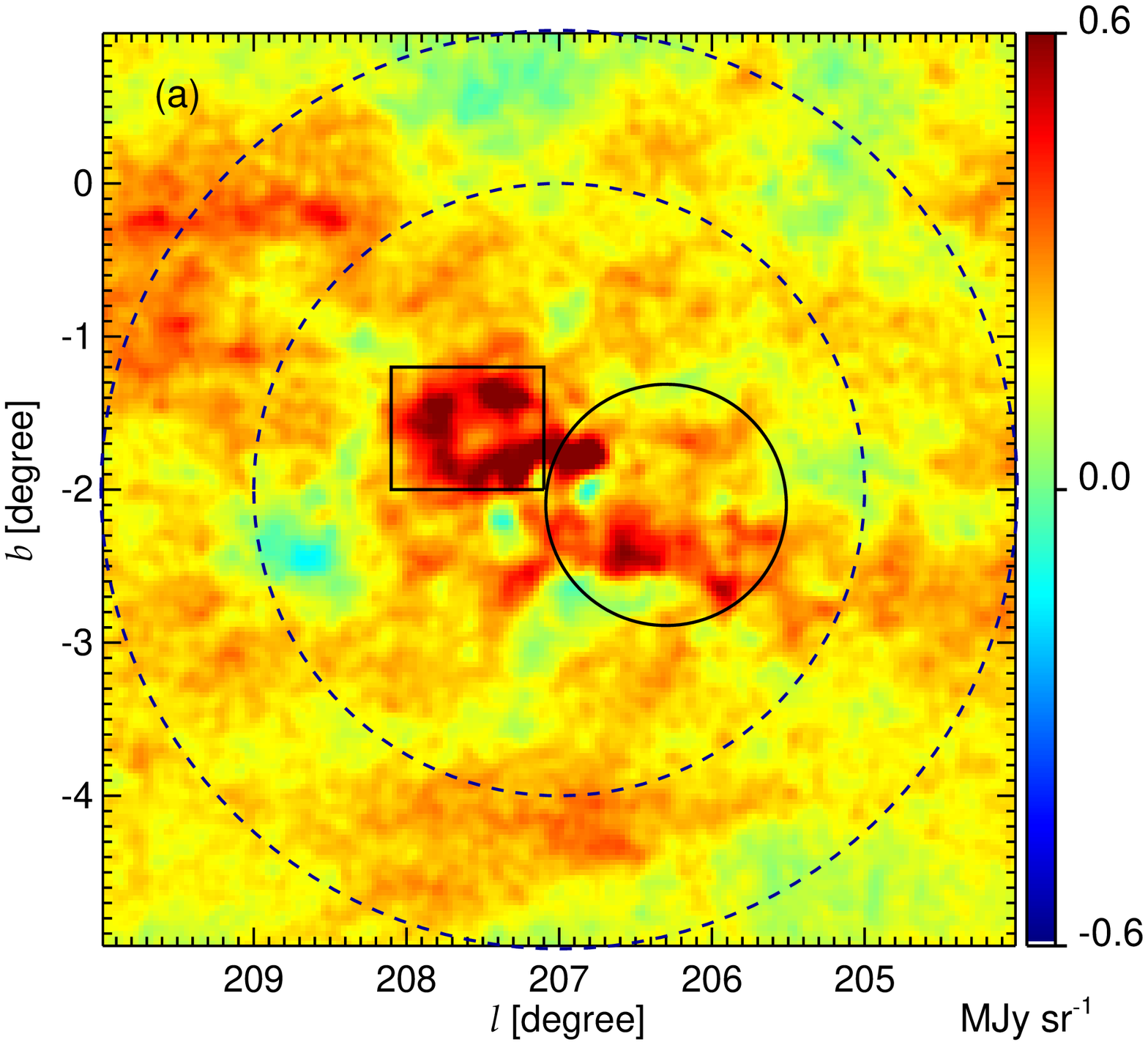}\vspace{-0.2cm}\hspace{0.2cm}
\includegraphics[scale=0.4]{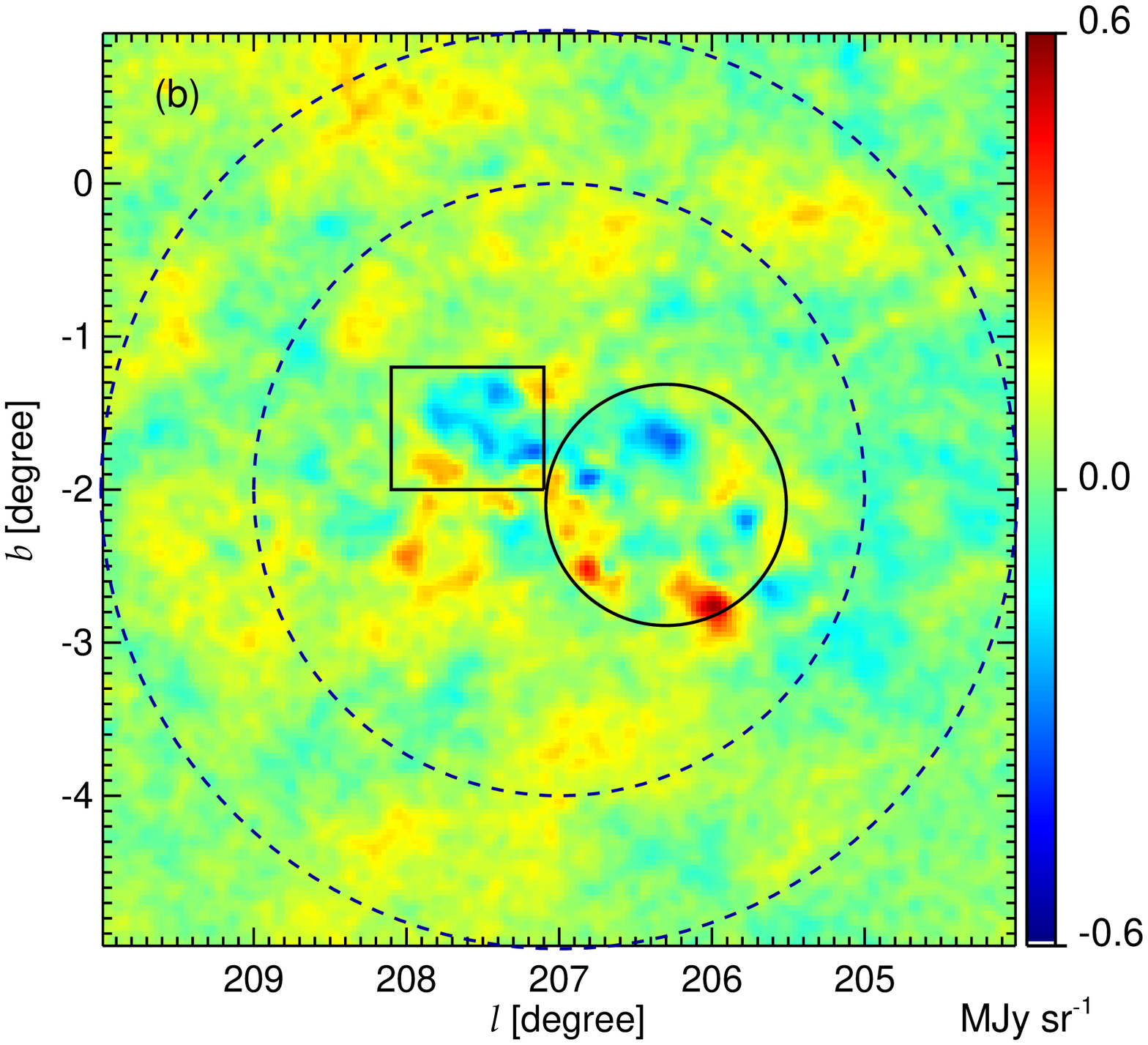}\vspace{-0.2cm}
\includegraphics[scale=0.4]{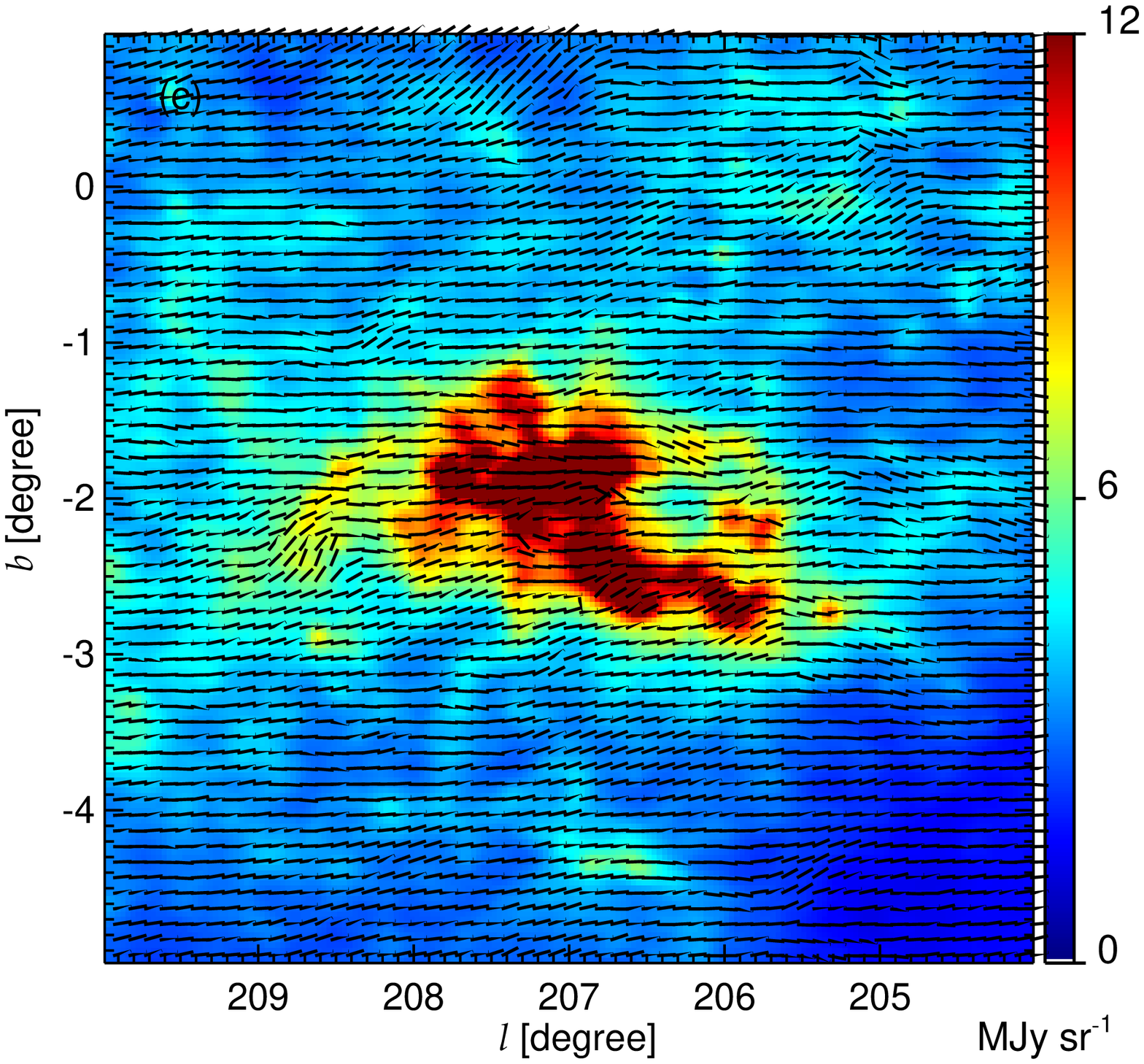}\hspace{0.2cm}
\includegraphics[scale=0.4]{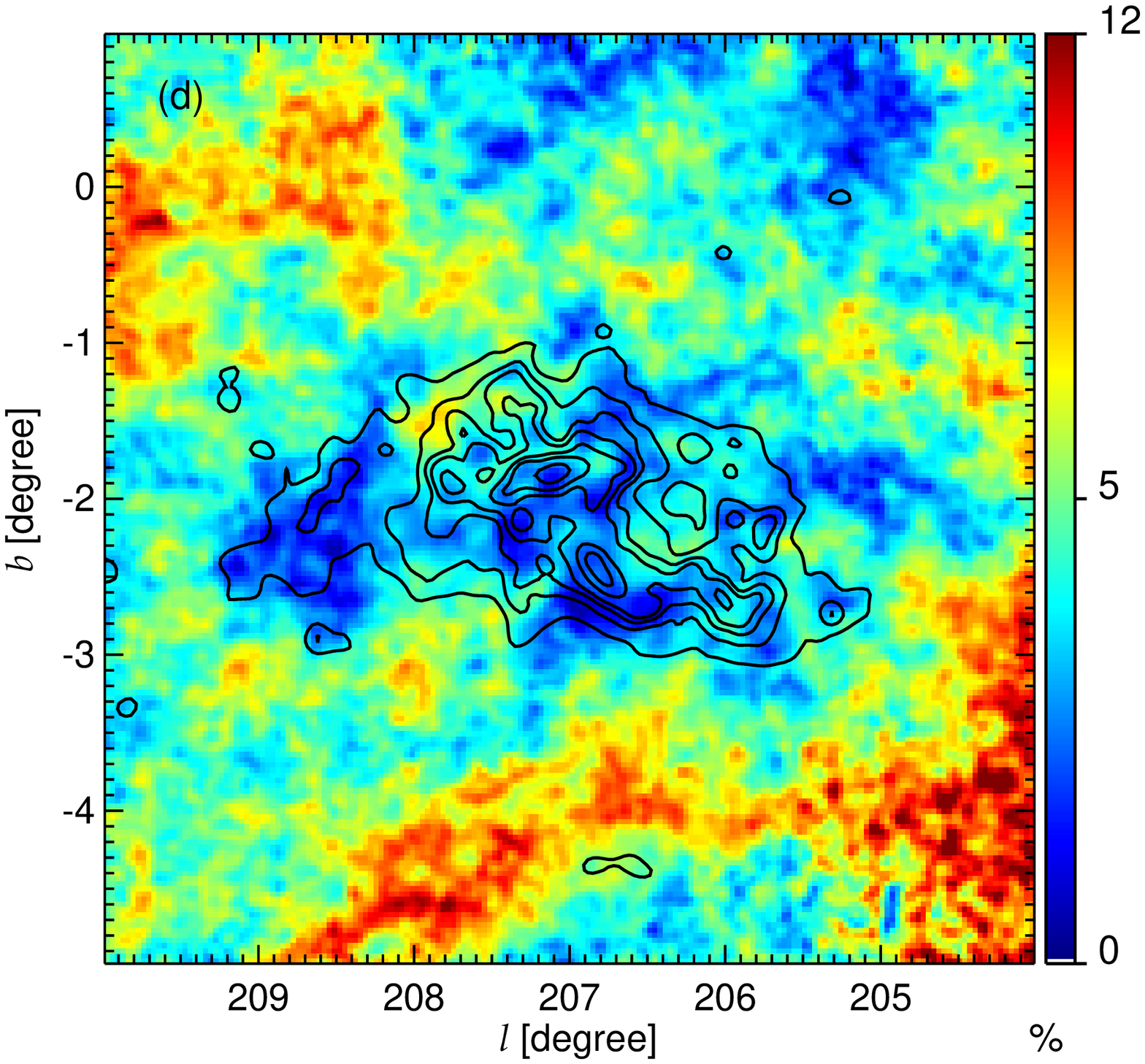}
\caption{\Planck\ polarization maps of the Rosette
region: (a) $Q$ Stokes parameter; (b) $U$ Stokes parameter. The dashed
circles, with radii 2\degr\ and 3\degr, define the region where we
estimate the background level, the
full circle centred on the star cluster represents the outer radius of
the dust shell, $r_{\rm out}^{\rm dust}=22$\,pc, and the rectangle
encompasses the brightest part of the 
Mon OB2 cloud. (c) Intensity map at 353\,GHz with the plane-of-the-sky magnetic field
orientation shown by headless vectors,
obtained by rotating the polarization angle $\psi$ by 90\degr. The
vectors are plotted at every 6\arcm, from the average of $Q$ and $U$
within the same distance from the central pixel. (d) Polarization fraction
$p$, with the same intensity contours as in Fig. \ref{fig:map1}.}
\label{fig:map2}
\end{figure*}

The western part of the Rosette complex has not been studied as thoroughly as the main
molecular cloud. It is not as bright in dust emission, but still visible in the \Planck\
and IRAS maps (Figs. \ref{fig:map1}(b), (e), and (f)) as a fragmented shell
surrounding the central cavity. 
There is no significant counterpart in CO emission, 
except in the cloud located in the south-west,
which appears elongated perpendicular to the ionization
front in the \Planck\ and CO maps. This cloud also has lower dust
temperature than the remainder of the western dust shell.

The dust shell seen in the \Planck\ 353\,GHz map surrounding the
\hii~region is associated with the Rosette complex, as it correlates with the
dust emission observed by IRAS \citep{Cox:1990} and identified in the dust
temperature map. The map of Fig. \ref{fig:map1}(d) shows a clear enhancement of the
dust temperature in the region occupied by the
nebula as a result of dust heating by the central
cluster. It also illustrates a decrease in the
dust temperature towards the central cavity because the Rosette is not a
filled \hii~region but a shell, with an inner cavity created by the
evacuation of material due to powerful stellar winds \citep{Smith:1973,Fountain:1979}.

The origin of the dust
shell as material swept up by the expanding \hii~region will be further
discussed in Sect. \ref{sec:magfield}. The size of the shell can
be estimated by fitting the \Planck\ intensity map with a shell model, similar to that used by
\citet{Celnik:1985} to derive the radius of the ionized nebula. We
obtain $r_{\rm in}^{\rm dust}=18$\,pc and $r_{\rm out}^{\rm dust}=22$\,pc for the inner and outer
radii of the dust shell, with a typical uncertainty of 2\,pc. 
Further details are given in Appendix \ref{appb}.

\subsection{Polarization}
\label{sec:pol}

\begin{figure*}
\centering
\includegraphics[scale=0.5]{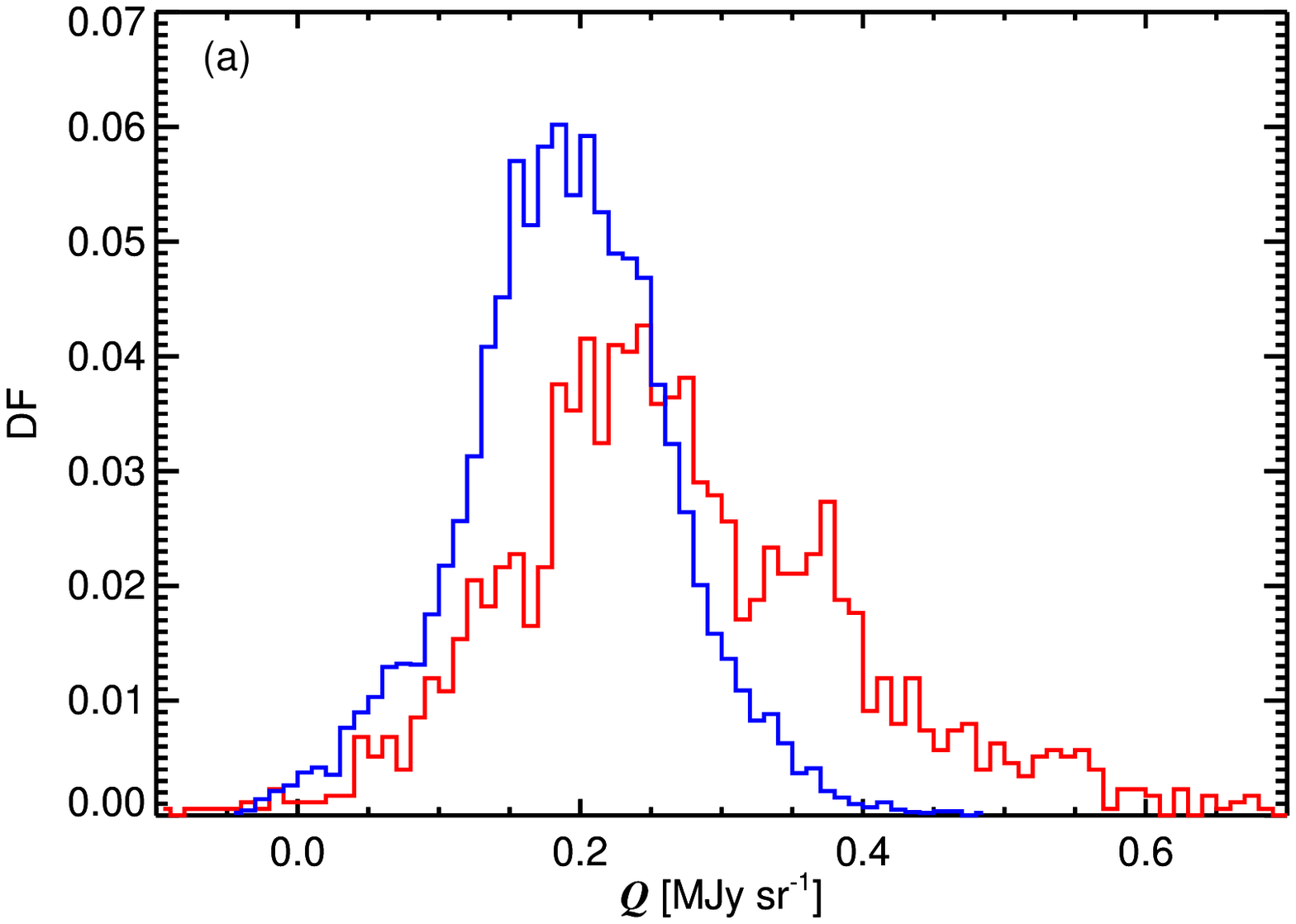}
\includegraphics[scale=0.5]{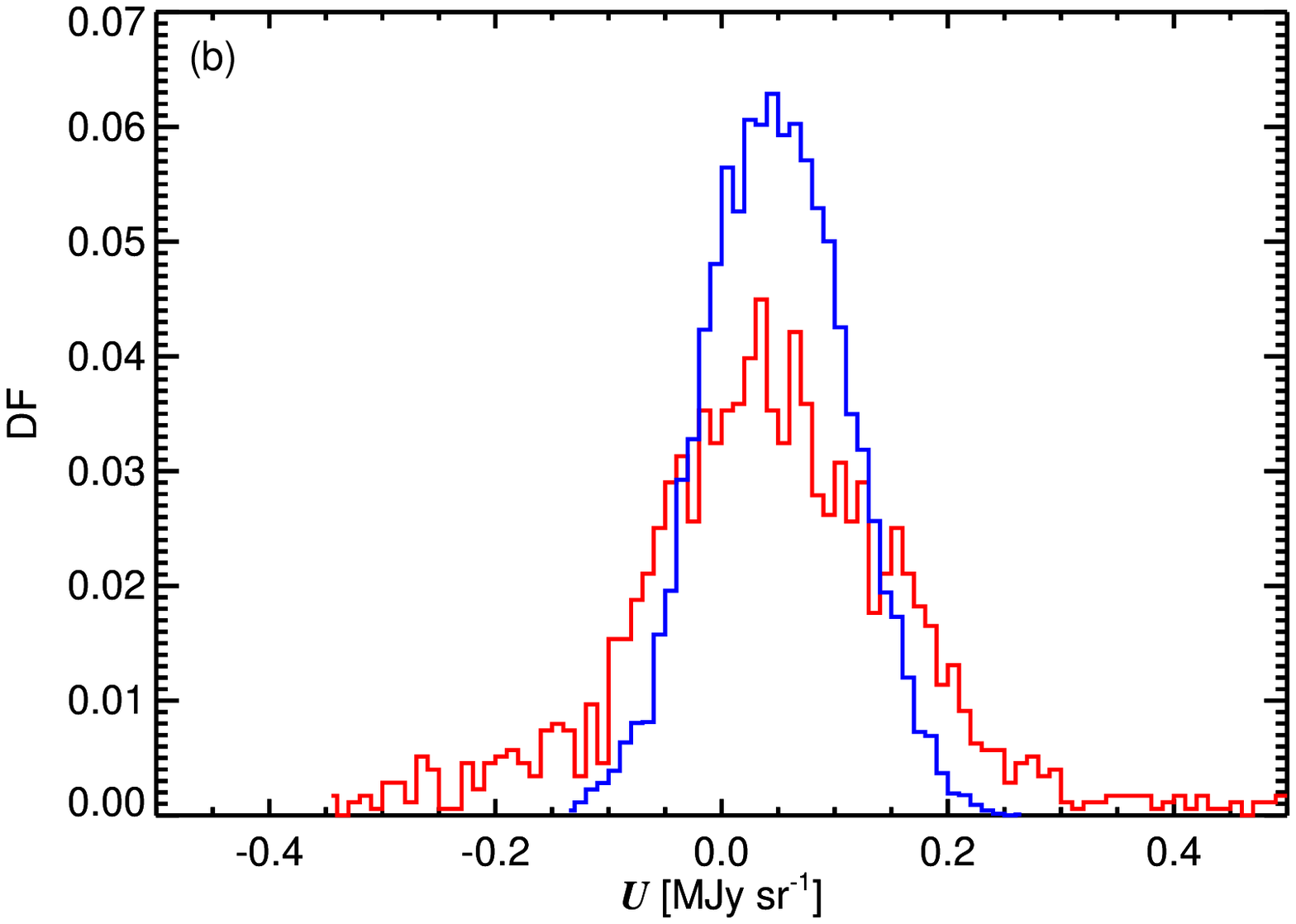}
\caption{Normalized distribution functions (DFs) of the (a) Stokes $Q$
   and (b) $U$ maps
  measured within the background aperture (blue) and within the outer
radius of the dust shell (red). The two regions are defined in
Fig. \ref{fig:map2} by the dashed and full circles, respectively.}
\label{fig:quhists}
\end{figure*}

\begin{table*}[tmb]
\begingroup
\newdimen\tblskip \tblskip=5pt
\caption{The mean level of the Stokes $I$, $Q$, and $U$
  parameters at 353\,GHz, in units of MJy\,sr$^{-1}$. The first three columns correspond
to the dust shell, the Mon OB2 cloud and the background regions,
respectively, as defined in the text and in Fig. \ref{fig:map2}. Their
uncertainties are given by the standard error on the mean.
The last two columns list the mean Stokes parameters in the dust shell
and Mon OB2 cloud, respectively, corrected for the background
contribution. The background subtraction is accounted for in the
uncertainties as described in Sect. \ref{sec:polplanck}.}
\label{table1}
\nointerlineskip
\vskip -3mm
\footnotesize
\def\leaderfi1{\leaders\hbox to 5pt{\hss.\hss}\hfil}
\setbox\tablebox=\vbox{
   \newdimen\digitwidth 
   \setbox0=\hbox{\rm 0} 
   \digitwidth=\wd0 
   \catcode`*=\active 
   \def*{\kern\digitwidth}
   \newdimen\signwidth 
   \setbox0=\hbox{+} 
   \signwidth=\wd0 
   \catcode`!=\active 
   \def!{\kern\signwidth}
   \halign{\hbox to 0.7in{#\leaderfil}\tabskip 8pt&
     \hfil#\hfil\tabskip 8pt&
     \hfil#\hfil\tabskip 8pt&
     \hfil#\hfil\tabskip 8pt&
     \hfil#\hfil\tabskip 8pt&
     \hfil#\hfil\tabskip 8pt&
     \hfil#\hfil\tabskip 0pt&
  \hfil#\hfil\tabskip 0pt\cr
     \noalign{\doubleline\vskip1pt}
 \omit\hfil Region \hfil & Dust shell & Mon OB2 & Background & Difference: Shell
 & Difference: Mon OB2 \hfil\cr 
\noalign{\vskip 3pt\hrule\vskip 5pt}
$\langle I \rangle$ & $9.44\pm0.08$& $\phantom{-}10.6\pm0.2$ & $3.631\pm0.006$ & $\phantom{-}5.80\pm0.09$ & $\phantom{-}7.02\pm0.2\phantom{0}$\cr
$\langle Q \rangle$ & $0.278\pm0.003$ & $\phantom{-}0.461\pm0.005$&$0.1911\pm0.0006$&$\phantom{-}0.087\pm0.004$ & $\phantom{-}0.270\pm0.005$ \cr
$\langle U \rangle$ & $0.045\pm0.003$ & $-0.004\pm0.005$&$0.0488\pm0.0005$ &$-0.004\pm0.004$ & $-0.053\pm0.005$\cr
\noalign{\vskip 5pt\hrule\vskip 3pt}}}
\endPlancktablewide
\endgroup
\end{table*}

\subsubsection{\textit{Planck} data}
\label{sec:polplanck}

The maps of the $Q$ and $U$ Stokes parameters towards the Rosette
Nebula and its parent molecular cloud are shown in
Figs. \ref{fig:map2}(a) and (b), respectively. 
There is significant $Q$ emission towards Mon OB2 relative to the
background, however the polarized emission surrounding the
Rosette \hii~region does not define the same shell as seen in
intensity. 
Notably, the $Q$ and $U$ signal in
some regions of the dust shell is at the same level as that outside the
Rosette/Mon OB complex. Figure \ref{fig:quhists} compares the $Q$ and
$U$ emission observed in the dust shell that surrounds the Rosette
\hii~region (in red) with that from an aperture that represents the local
background (in blue). The first distribution is measured within 22\,pc (the
outer radius of the dust shell, Sect. \ref{sec:int}) from the central
cluster and the second within an aperture of inner and outer radii of
2\degr\ and 3\degr, respectively, centred at
$(l,b)=(207\degr,-2\degr)$ (see the circles in
Fig. \ref{fig:map2}). 
As seen in Fig. \ref{fig:quhists}(a), the $Q$ signal in the shell
clearly deviates from that measured in the
background, with a higher mean value and a broader distribution.  
The distribution of $U$ emission, shown in Fig. \ref{fig:quhists}(b),
is also broader in the dust shell, even
if the mean level is similar to that of the background. We compare
the histograms of Fig. \ref{fig:quhists} quantitatively by means of the
Kolmogorov-Smirnov statistical test, which shows that the polarized
signal from the shell and the background are significantly
different. This test compares the cumulative distributions and thus
takes into account the different number of points from each sample.
The statistical difference applies when considering the dust shell as a whole, but
it may not be valid in the case of individual structures located
inside it. We will thus focus the analysis on the mean polarized
quantities, without attempting to study the small-scale structure
observed in the dust shell. This approach is also justified because
our analytical magnetic field model is based on a uniform
density distribution.

We estimate the mean value of the $I$, $Q$, and $U$ Stokes parameters
in the shell and background regions. These are listed in Table
\ref{table1} along with the corresponding uncertainties, given by the
standard error on the mean. The second to last column of Table~\ref{table1}
lists the mean of the Stokes parameters in the dust shell corrected
for the background emission, required for comparison with the model
results. We note that the choice of a constant value for the
background follows from the absence of any clear pattern in the
polarized emission outside the Rosette/Mon OB2 that we can use
to model the background variations. As a result, the
contribution of the background residuals to the polarized signal of
the dust shell needs to be taken into account. The background
distributions of Fig. \ref{fig:quhists} have about 8 times
more points ($N_{\rm back}=14\,154$) than those corresponding to the dust shell
($N_{\rm dust}=1\,757$). We extract $N_{\rm dust}$ points 
 from the background distribution and assess the variation of its mean Stokes
parameters by repeating this exercise 10\,000 times. We take the standard
deviation of the 10\,000 mean values and add it in quadrature
  to the standard error on the mean Stokes parameters. The final
  uncertainty in the background-subtracted values is quoted in the
  second to last column of Table \ref{table1}. We also computed these quantities using a
  different leakage correction (as explained in Sect. \ref{sec:planckd1}), which we will use in
Sect. \ref{sec:modeldust} to quantify the systematic uncertainties in our results. We note that the method
  used to compute the uncertainties does not take into account that
  the background fluctuations are spatially correlated. If we estimate the
background in circular areas of the same radius as the Rosette dust shell, we find a
larger uncertainty on the background-corrected values. Thus, we cannot
exclude that the mean $Q$ and $U$ values in Table \ref{table1} for the
dust shell have a significant contribution from local background
fluctuations. However, we regard the possibility of having a second
ISM structure in the same direction as the Rosette as
unlikely. 

We perform the same analysis in the rectangular region of Fig. \ref{fig:map2} that
delineates the brightest part of Mon OB2. The mean Stokes parameters, 
corrected for the background contribution are listed in the last
column of Table \ref{table1}. We use these values to estimate the polarization properties of
the Rosette's parent molecular cloud. The resulting debiased
polarization fraction, derived according to the method
  introduced in Sect. \ref{sec:planckd1}, is $p=(3.8\pm0.9)$\,\% and the polarization
angle is $\psi=-3\degr\pm14\degr$. The plane-of-the-sky magnetic field
orientation in Mon OB2, along the Galactic plane, is consistent with the large-scale orientation
seen in Fig. \ref{fig:map2}(c). 

The map of
polarization fraction is shown in Fig. \ref{fig:map2}(d), where the
lowest values of $p \lsim 3$\,\% are seen towards the densest regions surrounding
the Rosette Nebula. Similar values are observed in the less 
bright part of the Mon OB2 cloud at $(l,b)=(208\pdeg5,-2\pdeg5)$.
From the all-sky analysis of \Planck\ 353\,GHz polarization data,
\citet{planck2014-XIX} find that higher density regions have lower
polarization fraction relative to the maximum $p_{\rm max}=19.8$\,\%,
detected at 1\degr\ resolution.
This decrease in $p$
can be the result of several effects, namely depolarization in the beam or
along the line of sight, variations in the intrinsic
polarization fraction of dust grains, as well as changes in the
magnetic field geometry. The latter was shown to generally
explain the variations of $p$ across the
whole sky: \citet{planck2014-XIX} find an
anti-correlation between the fluctuations in the magnetic field
orientation and the polarization fraction.
\citet{planck2014-XXXIII} used this result to model the variations of the Stokes
parameters across three interstellar filaments with variations of the
magnetic field orientation, for a fixed alignment efficiency of dust grains.
In Sect. \ref{sec:magfield} we present an analytical solution for the magnetic field in the Rosette
\hii~region and associated dust shell, for constant dust properties. We will use this model to
explain the present \Planck\ polarization observations and the
radio RM data consistently.

\subsubsection{Radio data}
\label{sec:polrm}

The radio emission map of Fig. \ref{fig:map1}(a) shows the Rosette \hii~region and its
nearly circular shape. We fit the radial profile of the RRL emission
with a shell model and find the same values as \citet{Celnik:1985} for the inner and outer
radii, $r_{\rm in}^{\hii}=7$\,pc and $r_{\rm out}^{\hii}=19$\,pc, respectively,
with an uncertainty of 1\,pc. Also shown in Fig. \ref{fig:map1}(a) are
the positions of the RM data from \citet{Savage:2013}
(Sect. \ref{sec:rmd}). The circles have a diameter equivalent to the beam FWHM of the
radio survey, $14\farcm4$, in order to show the regions within which
the electron density, $n_{\rm e}$, is estimated.

The RRL observations provide a measure of EM for a given electron
temperature. We use Eq. (\ref{emeq2}) with $T_{\rm e}=8500$\,K
(Sect. \ref{sec:intro2}, \citealt{Quireza:2006}) to calculate EM
towards the 20 RM positions. The observed EM includes the
contribution from the warm ionized gas from
the background/foreground material in the Galaxy. We correct for this
contribution by subtracting the average EM measured towards the RM
positions located outside the \hii~region. Of the 20 positions, 14 lie
outside the radius $r_{\rm out}^{\hii}=19$\,pc, with an
  average EM of ($288\pm124$)\,cm$^{-6}$\,pc, where the uncertainty
  corresponds to the standard error on the mean. The EM values
  measured towards the shell are in the range 1500--5300
  \,cm$^{-6}$\,pc. We can thus
estimate the electron density in the Rosette towards the remaining six
positions, which are listed in Table \ref{table2}.
In the general case when the electron density distribution is not
uniform but concentrated in discrete clumps of ionized gas, 
the filling factor $f$ is introduced to relate the true path length $L$ of a
given line of sight through the nebula to the effective path length
$L_{\rm eff} = fL$, which is the total
length occupied by the individual clumps. The true path
length through a shell of inner and outer radii $r_{\rm in}^{\hii}$ and $r_{\rm out}^{\hii}$,
respectively, is given by \citep{Savage:2013}
\[
 L(\xi) =
  \begin{cases}
   2 r_{\rm out}^{\hii} \sqrt{1-\left(\xi/r_{\rm
          out}^{\hii}\right)^{2}} & \text{if } \xi \geq r_{\rm in}^{\hii} \\
   2 r_{\rm out}^{\hii} \left[\sqrt{1-\left(\xi/r_{\rm
           out}^{\hii}\right)^{2}} \right.  & \\
-  \left. \left(r_{\rm in}^{\hii}/r_{\rm out}^{\hii}\right)
  \sqrt{1-\left(\xi/r_{\rm in}^{\hii}\right)^{2}} \right]  & \text{if
} \xi < r_{\rm in}^{\hii} 
  \end{cases}
\]
where $\xi$ is the linear distance between a given line of sight and
the line of sight to the centre of the nebula, measured in the
transverse plane through the nebula. For the Rosette, with $r_{\rm
  in}^{\hii}=7$\,pc and $r_{\rm out}^{\hii}=19$\,pc as derived above,
$L(\xi=0)=24$\,pc is the path length through the 
centre of the shell.

\begin{table}[tmb]
\begingroup
\newdimen\tblskip \tblskip=5pt
\caption{Results from the analysis of the RM and EM data: Column 1 gives the
  RM source number, as in \citet{Savage:2013};
  Columns 2 and 3 list the Galactic coordinates of the source; Column
  4 gives the linear distance from the centre of the Rosette; Column 5
  lists the mean electron density estimated with Eq. (\ref{meanne}) for
  $f=1$; Column 6
  gives the RM data corrected for the background contribution; and
  the last column gives the line-of-sight
  component of the magnetic field estimated with Eq. (\ref{bparrm})
  for $f=1$. The uncertainties in both
  $\langle n_{\rm e} \rangle$ and $B_{||}$ are statistical, and thus do
  not include the systematic uncertainties involved, namely those on 
  the distance to the Rosette, its electron temperature, path length
  (which depends on the radii), electron density distribution, and the
  calibration uncertainty of the EM data. }
\label{table2}
\nointerlineskip
\vskip -3mm
\footnotesize
\def\leaderfi1{\leaders\hbox to 5pt{\hss.\hss}\hfil}
\setbox\tablebox=\vbox{
   \newdimen\digitwidth 
   \setbox0=\hbox{\rm 0} 
   \digitwidth=\wd0 
   \catcode`*=\active 
   \def*{\kern\digitwidth}
   \newdimen\signwidth 
   \setbox0=\hbox{+} 
   \signwidth=\wd0 
   \catcode`!=\active 
   \def!{\kern\signwidth}
   \halign{\hbox to 0.5in{#\leaderfil}\tabskip 8pt&
     \hfil#\hfil\tabskip 8pt&
     \hfil#\hfil\tabskip 8pt&
     \hfil#\hfil\tabskip 8pt&
     \hfil#\hfil\tabskip 8pt&
     \hfil#\hfil\tabskip 8pt&
     \hfil#\hfil\tabskip 0pt&
   \hfil#\hfil\tabskip 0pt\cr
     \noalign{\doubleline\vskip1pt}
 \omit\hfil Source \hfil & $l$ & $b$ & $\xi$
& $\langle n_{\rm e} \rangle$ & RM  & $B_{||}$\hfil\cr
 \omit \hfil & [\degr] & [\degr] & [pc] &
[cm$^{-3}$] &[rad\,m$^{-2}$] & [$\mu$G]\hfil\cr
\noalign{\vskip 3pt\hrule\vskip 5pt}
I6 & 205.7 & $-2.1$& 17.0 & $\phantom{0}9 \pm2$ &$676 \pm 68$ & $5 \pm 1$\cr
I7  & 206.2 & $-2.1$& \phantom{0}4.6  & $12 \pm 1$& $594 \pm 30$  & $2.3 \pm0.2$\cr 
I8  &  206.8 & $-2.4$ & 16.0  & $12 \pm 1$ &$\phantom{0}219 \pm 113$ &$1.1\pm 0.6$\cr
I10 & 205.9 & $-1.7$& 16.4  & $10 \pm 1$ &$\phantom{0}709 \pm 117$& $5\pm1$\cr
I12  & 206.6 & $-1.9$& \phantom{0} 9.6  & $13 \pm 1$ & $703 \pm 26$ & $2.1 \pm0.1$ \cr
I15& 206.3 & $-1.5$ & 15.9  & $\phantom{0}8 \pm 1$ &$501 \pm 24$ &  $4\pm 1$\cr
\noalign{\vskip 5pt\hrule\vskip 3pt}}}
\endPlancktablewide
\endgroup
\end{table}

The average electron density along the line of sight is estimated
from the EM observations using Eq. (\ref{emeq1}) as
\begin{equation}
\label{meanne}
\langle n_{\rm e} \rangle = \sqrt{\frac{{\rm EM} f}{L}}. 
\end{equation}
Table \ref{table2} lists the average electron density at each of the
six positions in the Rosette for $f=1$. The derived values
are between 8 and 13\,cm$^{-3}$ and are consistent with those obtained
by \citet{Celnik:1985}. The mean electron density across the
  entire \hii~region is 12.3\,cm$^{-3}$, with a scatter of
  4\,cm$^{-3}$. 

Similarly to the EM data, the measured values of RM include the contribution from the
large-scale magnetic field weighted by the Galactic warm ionized
gas. This needs to be corrected for in order to study the local
component of the field in the Rosette.
The mean of the 14 measurements
outside the radius $r_{\rm out}^{\hii}=19$\,pc is 
$(132\pm20)$\,rad\,m$^{-2}$, where the uncertainty
corresponds to the standard error on the mean. The background
measurements vary between 50 and 230\,rad\,m$^{-2}$ (half of these have an
uncertainty of more than 20\,\%). The final six background-corrected RM values
are listed in Table \ref{table2}.

The line-of-sight component of the magnetic field can be obtained by
combining the EM and RM observations, using Eqs. (\ref{rmeq}) and
(\ref{emeq1}), as
\begin{equation}
\label{bparrm}
B_{||} = \frac{ {\rm RM}/{\rm rad\,m^{-2}} }{ 0.81 \sqrt{ ({\rm EM}/{\rm
        cm^{-6}\,pc}) ( L/{\rm pc} ) f}}\, \mu{\rm G}.
\end{equation}
This approximation holds if $B_{||}$ is uniform across the
nebula and if $n_{\rm e}$ and $B_{||}$ are
uncorrelated along the line of sight. \citet{Beck:2003} point out that
the latter assumption can lead to uncertainties by a factor of 2--3 in the
estimated value of $B_{||}$. The derived $B_{||}$ values in
the nebula for $f=1$ are listed in Table \ref{table2} and vary from $+1$ to $+5\,\mu$G.
If we consider a filling factor $f=0.1$
\citep{Herter:1982,Kassim:1989}, the values increase by a factor of 3.2,
varying between $+3$ and $+16\,\mu$G. These results are in the range of
the $B_{||}$ values measured in the diffuse ISM
\citep{Crutcher:2012}. As will be discussed in the next
section, the hypothesis that the magnetic field is uniform throughout the
Rosette \hii~region is not satisfied. The field is confined to the
ionized shell and thus its direction must vary across the nebula. We
will assess this aspect by means of a 2D magnetic field
model and compare its predictions with the measured $B_{||}$ values.

\section{The magnetic field in the Rosette}
\label{sec:magfield}

In this section we interpret the observations by comparing them
with the results from an analytical model of an ionized nebula
that has expanded in a uniform and magnetized medium and formed a
neutral shell of swept-up matter. The model is presented in
Sect. \ref{sec:model} and further detailed in Appendix
\ref{appb}. We use it to study the RM distribution 
across the nebula (Sect. \ref{sec:modelrm}), as well as to reproduce
the mean polarization of the dust shell observed by \Planck\
(Sect. \ref{sec:modeldust}). 

\subsection{The magnetized Str{\"o}mgren shell}
\label{sec:model}

The evolution of an expanding ionized nebula has been studied
numerically, both in uniform and turbulent magnetized media (e.g.
\citealt{Krumholz:2007}, \citealt{Arthur:2011}). 
As the \hii~region expands, the surrounding ISM is swept up into a shell around the central
stars. In accordance with the frozen-in condition, magnetic field lines are dragged with
the expanding gas and concentrated in the dense shell. 
If the magnetic pressure is
comparable to the thermal pressure in the \hii~region, 
magnetic forces lead to departures from sphericity. Furthermore, because
the swept-up magnetic flux increases from the
magnetic poles (along the direction of the initial field \vec{B}$_{0}$) to the
equator (90\degr\ from \vec{B}$_{0}$), magnetic
pressure in the swept-up shell
tends to make the shell thickness similarly increase from the poles to the equator.
We will ignore magnetic effects and assume that the \hii~region expands equally in all directions,
creating a spherical neutral shell of swept-up ISM. We consider
this assumption to be consistent with the radio observations of the
Rosette \hii~region, which show its nearly circular shape (e.g.
Fig. \ref{fig:map1}(a)), despite a possible elongation along
the line of sight.

Within this framework, we derive an analytical solution for the
magnetic field in a spherical structure composed of a shell of
swept-up gas formed around a shell of ionized gas. This configuration is shown in
Fig. \ref{fig:app0} of Appendix \ref{appb}, where the details of the
derivation are discussed.
The model is an analytical description of
the correspondence between the initial and present configurations of
the radial distribution of matter, which characterizes the expansion of the
\hii~region. We consider that the initial gas density and the magnetic field \vec{B}$_{0}$
are uniform and that they evolve as the matter expands radially.
This is translated in the form of an expansion law
(Fig. \ref{fig:app3}). The final magnetic field \vec{B}
(Eq. \ref{model7}) depends on the strength of the initial
field, $B_{0}$, and its direction, described by the polar and azimuthal
angles $(\theta_{0}, \phi_{0})$, with respect to the line of sight.

We note that the Galactic magnetic field has uniform and
  random components, which are of the same order (see
  e.g. \citealt{Beck:2001}). In the present study, \vec{B}$_{0}$
  corresponds to the initial local field, which is a sample of the
  total (uniform plus random) magnetic field.

In the following sections we describe how the combination of 
RM and dust polarization data allows us to fully describe the magnetic field
in the Rosette and its parent molecular cloud. These data have
different resolutions; however, this is not a concern since the model
does not attempt to reproduce any fluctuations on the scale of the
resolution of either of the observables.
\begin{figure}
\centering
\includegraphics[scale=0.5]{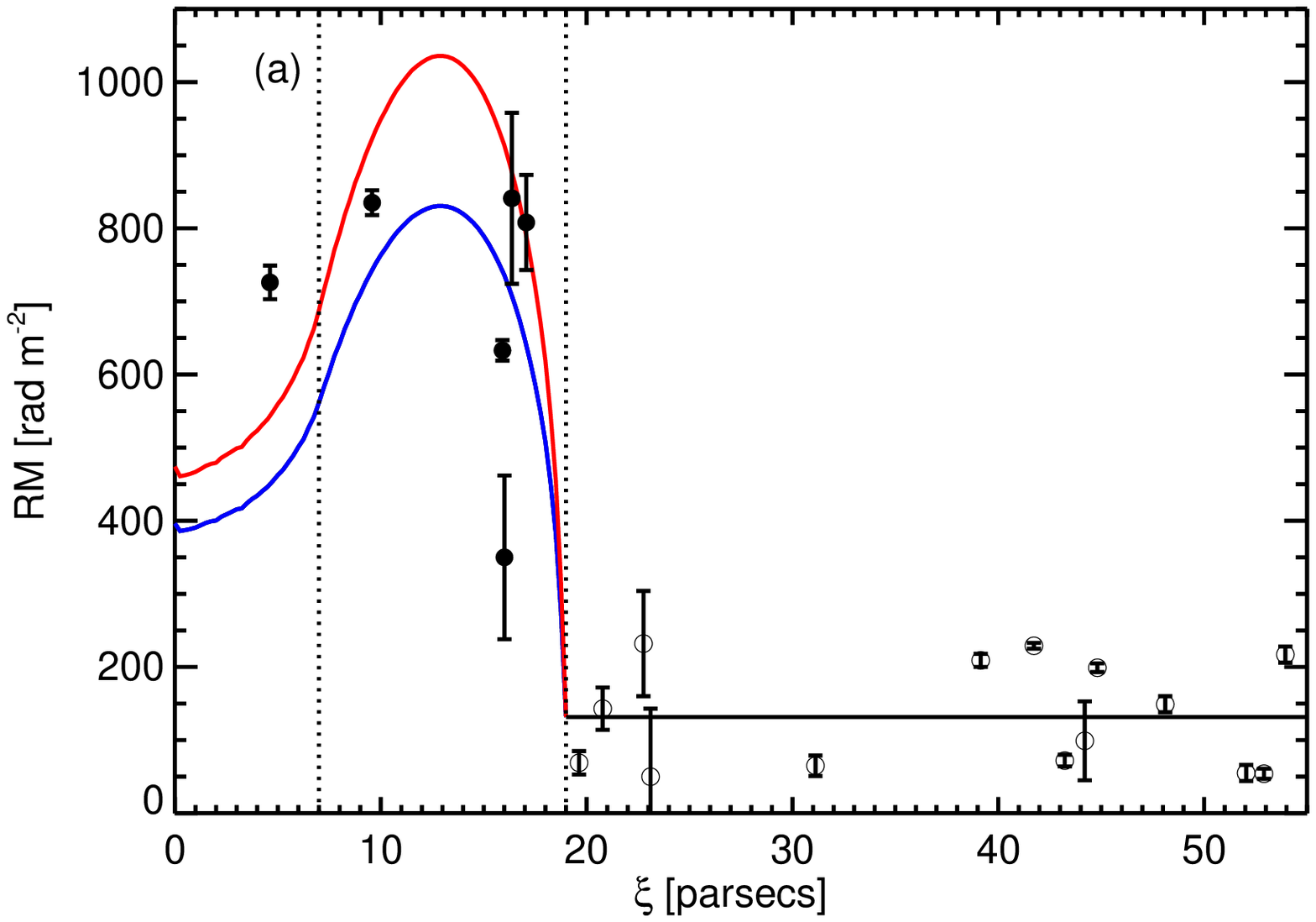}
\includegraphics[scale=0.5]{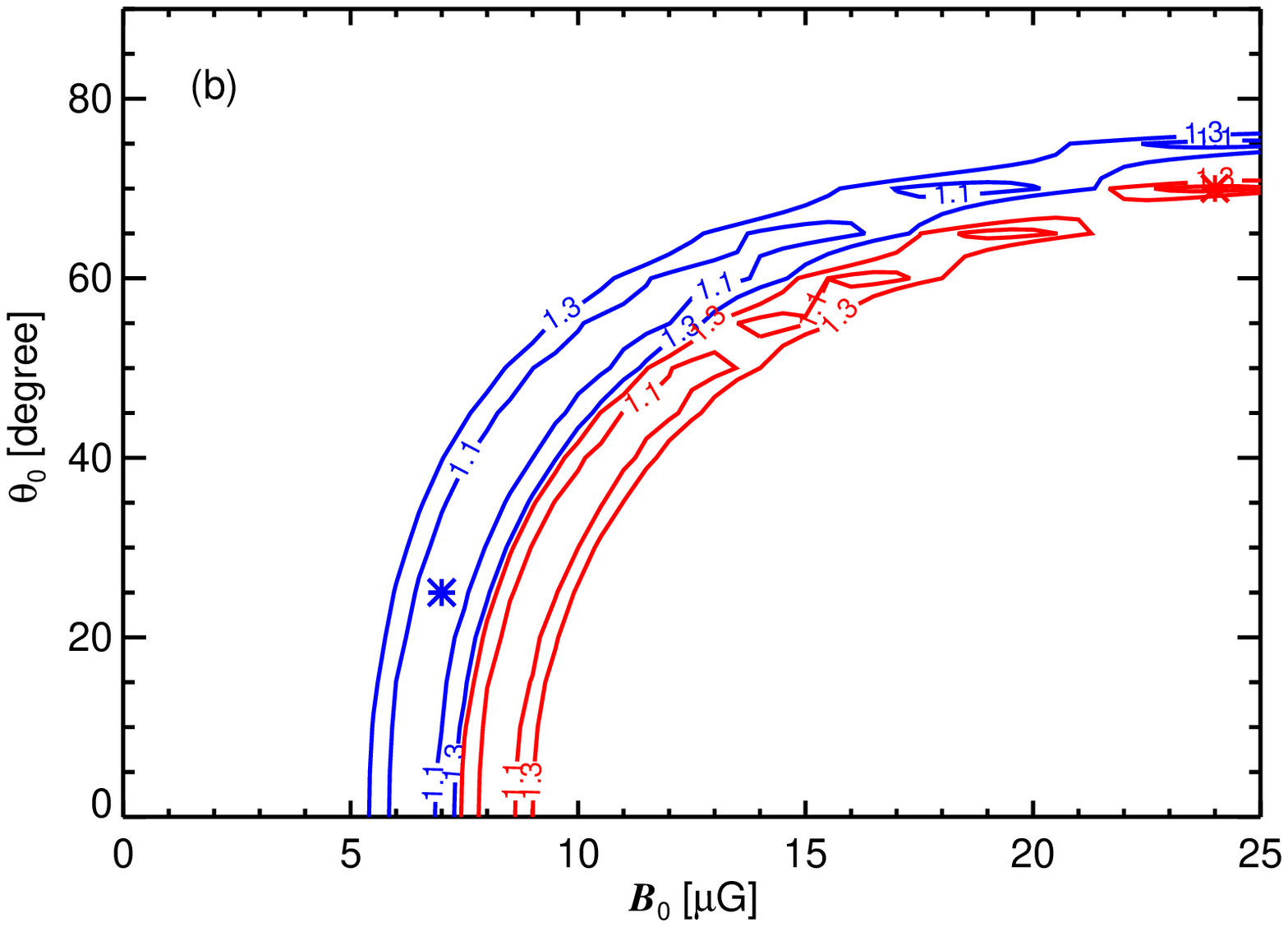}
\caption{Comparison between the RM observations and the predictions
  from the model. Panel (a) shows the RM data as a function of the
  distance to the centre of the Rosette. The two vertical dashed lines
  show the inner and outer radii of the nebula, $r_{\rm in}^{\hii}=7$\,pc and $r_{\rm
    out}^{\hii}=19$\,pc, respectively (Sect. \ref{sec:polrm}). The blue curve shows the radial
  profile of the modelled RM for the best fit to the six RM measurements within $r_{\rm
    out}^{\hii}$ (filled circles), the reference fit. The red curve is the result of
  fitting the highest four RM data points. The open circles correspond
  to the RM observations outside the Rosette, used to estimate the
  background RM, and are not included in
  the fit. Panel (b) presents the reduced $\chi^{2}$ from both
  fits. The stars indicate the best fit parameters $B_{0}$ and
  $\theta_{0}$, which correspond to the minimum $\chi_{\rm r}^{2}$ for
  each fit. The reference fit (blue) gives
  $B_{0}=7\,\mu$G and $\theta_{0}=25\degr$ for $\chi^{2}=271$ and
  ${\rm N_{\rm dof}}=4$. The second fit (red) gives $B_{0}=24\,\mu$G and
  $\theta_{0}=70\degr$ for $\chi^{2}=76$ and ${\rm N_{\rm
      dof}}=2$. The contours are at 10 and
  30\,\% above the corresponding minimum values of $\chi_{\rm r}^{2}$. }
\label{fig:modelrm}
\end{figure}

\begin{figure}
\centering
\hspace{-0.3cm}
\includegraphics[scale=0.5]{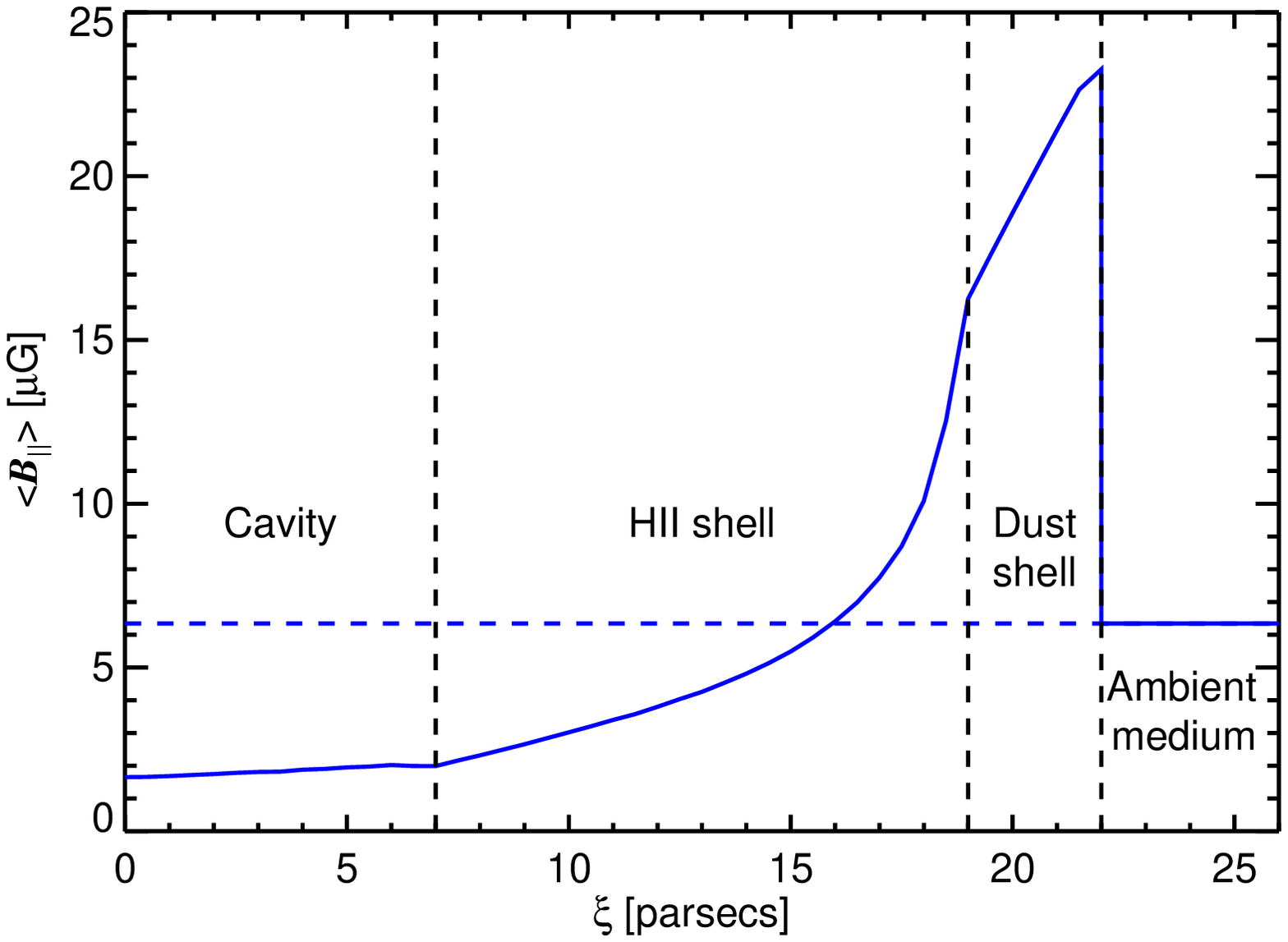}
\caption{Average of the line-of-sight component of the magnetic
  field as a function of the linear distance to the centre of the
  nebula. The curve shows the result of the reference fit, for which
  $B_{0}=7\,\mu$G, $\theta_{0}=25\degr$, and $B_{0||}=6.3\,\mu$G, as
indicated by the horizontal dashed line. 
  The three vertical lines delineate the
  inner and outer radii of the \hii~region and the outer radius of
  the dust shell: $r_{\rm in}^{\hii}=7$\,pc; $r_{\rm
    out}^{\hii}=19$\,pc; and $r_{\rm out}^{\rm dust}=22$\,pc. }
\label{fig:modelbpar}
\end{figure}

\subsection{Ionized shell: RM}
\label{sec:modelrm}

The RM of the modelled \hii~region is computed using
Eq. (\ref{rmeq}). We consider a uniform nebula with constant electron
density $n_{\rm e}=12.3$\,cm$^{-3}$, the mean value derived from the
EM data (Sect. \ref{sec:polrm}), and integrate $B_{||}$ along the
line-of-sight depth of the \hii~shell. Since the RM is derived solely from the line-of-sight
component of the field, it does not depend on $\phi_{0}$. 

We derive the
$B_{0}$ and $\theta_{0}$ values that best fit the RM data through a
$\chi^{2}$-minimization procedure
taking into account the corresponding uncertainties, as follows:
\begin{equation}
\label{fitchi2}
\chi^{2} = \sum_{N_{\rm points}} \frac{({\rm RM}_{\rm mod} - {\rm RM}_{\rm
    obs})^{2}}{\sigma_{\rm RM,obs}^{2}}.
\end{equation}
The reduced $\chi^{2}$ is computed as $\chi^{2}_{\rm r}=\chi^{2}/{\rm
  N_{dof}}$, where ${\rm N_{dof}}=N_{\rm points}-N_{\rm params}$ is the number of
degrees of freedom, the difference between the number of points and
the number of parameters in the fit. In the present case
  $N_{\rm params}=2$, corresponding to $B_{0}$ and $\theta_{0}$. The results are shown in
Fig. \ref{fig:modelrm}. Figure \ref{fig:modelrm} (a) presents the
best-fit RM curve as a function of the linear distance to the centre of the
nebula, $\xi$, along with the 20 RM measurements discussed in
Sect. \ref{sec:polrm}. The blue line is the 
result of a fit to the six RM data points within the outer radius of the
Rosette nebula, $\xi \le r_{\rm out}^{\hii}=19$\,pc, and hereafter
referred to as the reference fit. The best-fit
parameters are $B_{0}=7\,\mu$G and $\theta_{0}=25\degr$.
The shape of the RM curve is set by the radii of
the \hii~shell (thus by the expansion law used, Appendix \ref{appb}), and the scaling of the
curve depends on a combination of the two free
parameters. The model is unable to reproduce
the scatter in the RMs observed close to the boundary of the
\hii~region, which could be due to the possible clumpiness of the
medium and/or fluctuations in the magnetic field direction. These
effects are not accounted for in our model, which also assumes that
the expansion is perfectly spherical. 
We did not attempt to adjust the radii of the \hii~shell, or
equivalently its expansion law. Instead, we perform a second fit,
excluding the two lowest RM data points, at $\xi\sim16$\,pc, to
assess the departures of the observations from the simplified
assumptions of the present model. The result is shown by the red curve
in Fig. \ref{fig:modelrm} (a). The best-fit
parameters are $B_{0}=24\,\mu$G and $\theta_{0}=70\degr$, for
$\chi^{2}=76$ with ${\rm N_{\rm dof}}=2$, leading to a $\chi_{\rm
  r}^{2}$ that is about 30\,\% lower than that obtained in the
reference fit.

The $\chi^{2}$ contours as a function of $B_{0}$ and $\theta_{0}$ for the two
fits are shown in Fig. \ref{fig:modelrm} (b). 
The high $\chi^{2}$ values (given in the caption) are
  due both to the simplicity of the model and to the
uncertainties in the RM observations, which do not reflect their true radial
variation across the Rosette, as they correspond to individual
line-of-sight measurements.
The contours illustrate the degeneracy
between the strength and the orientation of the initial field relative
to the line of sight, as expected since ${\rm RM}\propto B_{||} = B
\cos\theta$ (Eq. \ref{rmeq}). As a consequence, all the
$(B_{0},\theta_{0})$ combinations that follow the minimum
$\chi^{2}_{\rm r}$ contour lead to approximately the same $B_{0||}$
value: $B_{0||}\simeq6\,\mu$G for the reference fit and
$B_{0||}\simeq8\,\mu$G for the second fit. Furthermore, for a given angle
$\theta_{0}$, the two different fits
give $B_{0}$ values that differ by less than about 4\,$\mu$G.

In the rest of this section, we consider only the reference fit obtained
when using all of the six RM measurements (blue curve in Fig. \ref{fig:modelrm}). 
Figure \ref{fig:modelbpar} shows the radial profile of the mean of
$B_{||}$, $\langle B_{||} \rangle$, measured
along the line of sight.
There is a significant difference between $\langle B_{||} \rangle$ in
the \hii~region and in the dust shell. We note that the exact shape of
the curve is determined by the adopted
expansion law (Fig. \ref{fig:app3}), which characterizes the two
distinct regimes in the evolution of the Rosette: the expansion of the
ionized gas, leading to a decrease in $B$ relative to $B_{0}$, and the
compression of the interstellar gas in the dust shell, accompanied by a compression
of the field lines, and hence an increase in $B$ (see Fig. \ref{fig:app4}).
The mean value of $B_{||}$ 
across the projected surface of the ionized shell, which in 3D includes the central cavity, is
2.6\,$\mu$G. This is comparable to the mean of the six values derived
from the RM data listed in Table \ref{table2}. Therefore, the present
model indicates that we can use the RM data to recover the mean
$B_{||}$ in the \hii~region, which is 62\,\% lower than $B_{0||}$ in
the molecular cloud.

The RM observations are from \citet{Savage:2013}, who fitted
a different analytical model to the data, as introduced in
Sect. \ref{sec:intro2}. The authors found a value of
$\theta_{0}=72\degr$ for an assumed $B_{0}=4$\,$\mu$G, under the assumption that a
strong adiabatic shock produces
an enhancement of the component of the field parallel to the expansion
front, relative to the ambient medium. 
In addition, \citet{Savage:2013} 
applied the shock boundary conditions to the whole thickness of the shell,
although these only hold in the thin-shell approximation. The present
magnetic field solution in Eq. (\ref{model7}) naturally explains the variations of 
the normal and tangential components of the field relative to the expansion
front, throughout the nebula.

This analysis defines the loci of $B_{0}$ and $\theta_{0}$ values
that best fit the RM data towards the Rosette nebula
(blue curve in Fig. \ref{fig:modelrm} (b)). 
For all of these solutions the resulting line-of-sight field
component in the ambient medium is $B_{0||}\simeq 6$\,$\mu$G, which
is at the low end of the range of values reported by \citet{Crutcher:2012} for
molecular clouds of similar column density as Mon OB2 (around
$3\times10^{22}$\,cm$^{-3}$,
\citealt{planck2013-p06b}). In the
following section we show that the degeneracy between $B_{0}$ and $\theta_{0}$
can be alleviated by further comparing the predictions from our model
with the \Planck\ polarization observations towards the dust shell.

\subsection{Neutral shell: dust polarized emission}
\label{sec:modeldust}
 
We model the shell that surrounds the \hii~region with constant
intrinsic dust polarization fraction $p_{0}$ and with inner and outer
radii of $r_{\rm in}^{\rm dust}=19$\,pc and $r_{\rm out}^{\rm
  dust}=22$\,pc, respectively (see Appendix \ref{appb}).
Since we will be comparing the ratios between the mean Stokes parameters,
we do not specify the density or temperature of the gas and work with 
normalized quantities. The polarization fraction $p$ can be written as
$p=p_{0}\sin^{2}\theta$.
We compute
$q=p_{0}\sin^{2}\theta\cos(2\psi)$ and
$u=p_{0}\sin^{2}\theta\sin(2\psi)$ (Eq. \ref{poleqs}) at every
position in the 3D shell and integrate along the line-of-sight
direction to obtain the normalized Stokes parameter maps.

\begin{figure}
\centering
\hspace{-0.2cm}
\includegraphics[scale=0.55]{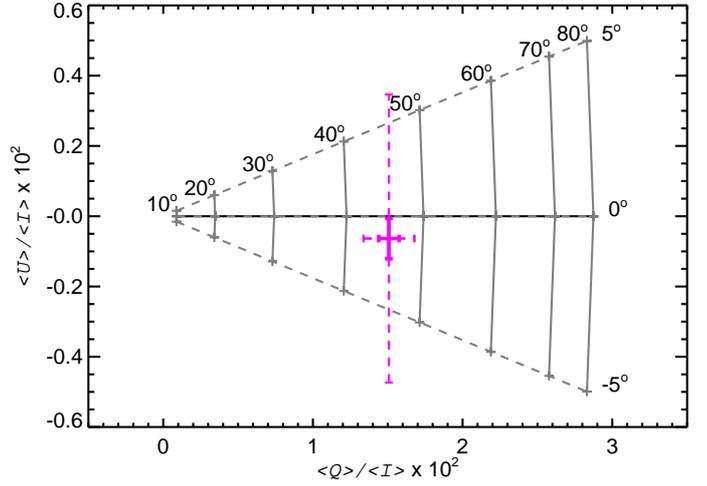}
\caption{Comparison between the \Planck\ observations and the
  model. The ratios $\langle Q \rangle / \langle I \rangle$
    and $\langle U \rangle / \langle I \rangle$ 
  derived from the polarization data are given by the magenta
  point. The statistical and systematic uncertainties are shown by the
  solid and dashed error
  bars, respectively. The dashed grey lines
  show the solutions for $\phi_{0}=-5\degr$, 0\degr, and 5\degr, with
  $\theta_{0}$ varying from 10\degr\ to 80\degr. 
  The solid grey lines indicate how the ratios change with $\phi_{0}$, for a given $\theta_{0}$ angle. 
  The model results are calculated here for $p_{0}=4$\,\%.}
\label{fig:polmod1}
\end{figure}

\begin{figure}
\centering
\hspace{-0.2cm}
\includegraphics[scale=0.55]{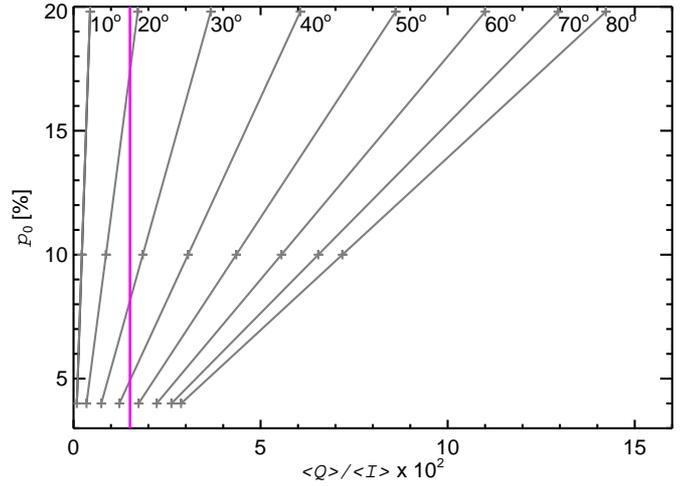}
\caption{Variation of the modelled $\langle Q \rangle / \langle I \rangle$ ratio as a function 
of $p_{0}$ and $\theta_{0}$, for $\phi_{0}=0\degr$ (grey lines). The
observed $\langle Q \rangle / \langle I \rangle$ 
ratio is shown by the vertical magenta line.}
\label{fig:polmod2}
\end{figure}

There are three variables involved in modelling the dust polarization:
the angles $(\theta_{0}, \phi_{0})$, which define the direction of
$\vec{B}_{0}$ with respect to the line of sight, and the intrinsic polarization
fraction $p_{0}$. The Stokes
parameters do not depend on the strength of the magnetic field. The polarization fraction
relation given above indicates that $p_{0}$ and $\theta$ are degenerate. We thus start by comparing
the data with the model
for a fixed $p_{0}$ value of 4\,\%, which corresponds to the observed
value in the Mon OB2 cloud (Sect. \ref{sec:polplanck}). Figure \ref{fig:polmod1} illustrates
how the predicted ratios between the mean values of the Stokes
parameters, 
$\langle Q \rangle / \langle I \rangle$ and $\langle U \rangle /
\langle I \rangle$, 
vary as a function of the initial magnetic field direction $(\theta_{0},\phi_{0})$.
The observed ratios, calculated using the values listed in the fifth column of
Table \ref{table1}, are $\langle Q \rangle / \langle I
\rangle = (1.51\pm0.07\,\rm{(sta.)
}\,\pm 0.17\,\rm{(sys.)})\times10^{-2}$
and $\langle U \rangle / \langle I \rangle = (-0.06\pm0.06\,\rm{(sta.)
}\,\pm 0.41\,\rm{(sys.)})\times10^{-2}$. The systematic uncertainties
correspond to the difference in the ratios when derived with the
different leakage correction maps (Sect. \ref{sec:planckd1}). The observed
$\langle U \rangle / \langle I \rangle$ ratio constrains the sky projected 
orientation of the initial magnetic field, $\phi_{0}\simeq0\degr$, to
within about 5\degr. This value is consistent with that 
measured towards the Rosette's parent molecular cloud
(Sect. \ref{sec:polplanck}) and corresponds to a magnetic field
parallel to the Galactic plane. Fixing this parameter 
allows us to study how the $\langle Q \rangle / \langle I \rangle$ ratio varies as a function
of the angle between \vec{B}$_{0}$ and the line of sight $\theta_{0}$,
and the intrinsic polarization fraction $p_{0}$. 
This is shown in Fig. \ref{fig:polmod2}, for $p_{0}$ ranging from 4 to 19.8\,\%, the maximum
polarization fraction observed across the sky
\citep{planck2014-XIX}. The comparison between the data and the
different models gives an upper limit on $\theta_{0}$ of about 45\degr\ for
$p_{0}=4$\,\%. A lower limit of $\theta_{0}\simeq 20\degr$, implying a
field that is nearly along the line of sight, is obtained for the
maximum intrinsic polarization fraction $p_{0}=19.8$\,\%.

The comparison between the present model and the \Planck\ polarization observations leads to
two main results. First, the inferred range of $\theta_{0}$,
combined with the results from the RM study, restricts the range of
the magnetic field strength in the Rosette's parent molecular cloud to
$B_{0}\simeq 6.5$--9\,$\mu$G. Second, we find that the Mon OB2 
 cloud has a magnetic field structure distinct from that of the Perseus spiral arm; an azimuthal
 Galactic mean field with a pitch angle of $-8\degr$ \citep{Ferriere:2011}, is expected to be
 oriented at about 60\degr\ from the line of sight at the position of Mon
 OB2. Within the uniform density and polarization fraction assumption
 of this model, such values of $\theta_{0}$ are only possible for a
 significantly low intrinsic 
 polarization fraction $p_{0}$. However, we cannot discard
 depolarization effects from a turbulent field and/or clumpy density
 distribution. While several
 observational studies indicate that molecular clouds preserve the large-scale field orientation
 (see \citealt{Li:2014}), \Planck\ observations question this general interpretation of 
 almost no variation of the magnetic fields with interstellar
 structures. Modelling of the dust polarization
 data from \Planck\ allows us to study the 3D geometry
 of the magnetic field in a variety
 of environments. The analysis of the magnetic field structure in
 nearby interstellar filaments by \citet{planck2014-XXXIII} 
 suggests that their evolution is coupled to the field, which is
 distinct from the field of the clouds in which they are
 embedded. This result agrees with our findings on the Rosette/Mon OB2 complex.

Finally, we can estimate the magnetic and thermal pressures in the
\hii\ and dust shells. We do not include the possible
  contribution from a turbulent field at scales much smaller than the
  Rosette, whose quantification
  is difficult. In either case this is expected to be small owing to the
  fast expansion of the \hii~region, which tends to order the magnetic
  field. For an initial field strength
$B_{0}=9\,\mu$G, which corresponds to $\theta_{0}=45\degr$, the
modelled magnetic field has a mean value $B^{\hii}=3.2$\,$\mu$G
within the ionized nebula and $B^{\rm dust}=21.4$\,$\mu$G in the
dust shell. The thermal pressure in the
\hii~region is $P_{\rm th}^{\hii}\simeq2n_{\rm e}kT_{\rm
  e}=2.9\times10^{-11}$\,erg\,cm$^{-3}$, where $k$ is
the Boltzmann constant, $n_{\rm e}=12.3$\,cm$^{-3}$, and $T_{\rm
  e}=8500$\,K (Sects. \ref{sec:intro2} and \ref{sec:polrm}). 
The magnetic pressure, $P_{\rm mag}^{\hii}=(B^{\hii})^{2}/(8\pi) =
0.4\times10^{-12}$\,erg\,cm$^{-3}$, is therefore smaller than the thermal pressure in the
\hii~region, supporting our initial assumption that the \hii~region is
roughly spherical.  In the dust shell, the magnetic pressure is $P_{\rm
  mag}^{\rm dust}=1.8\times10^{-11}$\,erg\,cm$^{-3}$, hence smaller than
but of the same order as the ionized gas pressure. This implies that both the
thickness and the density of the dense shell should not be constant
(e.g. \citealt{Ferriere:1991}). Modelling these effects is beyond the
simplified nature of the present analysis, where we focus on the mean
polarization properties of the Rosette, but will be taken into account
in future work. 

\section{Conclusions}
\label{sec:conc}

This work represents the first joint analysis and modelling of radio and submillimetre polarization 
observations towards a massive star-forming region (the Rosette Nebula) to study its 3D 
magnetic field geometry. We have developed an analytical solution for the magnetic field, 
assumed to evolve from an initially uniform configuration following the expansion of ionized gas
and the consequent concentration of the surrounding ISM in a dense
shell. The assumption of uniform density and temperature distributions for both the ionized and dust shells,
along with constant intrinsic polarization fraction of dust grains, is clearly an approximation. 
Different parts of the fragmented, swept-up shell presumably have
distinct properties and some may even
be pre-existing dense clouds caught by the expanding \hii~region. 
Nevertheless, the model
is able to reproduce the mean observed quantities.

We use the \Planck\ data at 353\,GHz to trace the dust emission from the shell of swept-up ISM
surrounding the Rosette \hii~region. Even if the shell is clearly seen
in intensity, the same pattern is not detected 
in dust polarized emission against the local background. When analysed
as a whole, the polarized signal from the dust shell is 
significantly distinct from that of the background and can be reproduced by the current model. 
The correspondence between the model and the \Planck\ observations
constrains the direction of the magnetic field in the Rosette's parent
molecular cloud Mon OB2 to an angle in the plane of the sky $\phi_{0}
\simeq0\degr$ (roughly parallel to the Galactic plane) and an angle to
the line of sight $\theta \lsim 45\degr$. This result is crucial to
removing the degeneracy between $\theta_{0}$ and 
$B_{0}$ inherent in the RM modelling. We thus find that $B_{0}$ is about  
6.5--9\,$\mu$G in Mon OB2.

The present magnetic field model provides a satisfactory fit to the
observed RM distribution as a function
of the distance from the centre of the Rosette \hii~region. More data are needed
to better understand the abrupt variations of RM close to the outer radius of the nebula. 
The RM modelling suggests a significant increase in the 
line-of-sight magnetic field from the \hii~region to the dense shell,
where $B_{||}$ reaches nearly 4 times $B_{0||}$.

The combination of RM and dust polarization data in this work is essential to constrain both
the direction and the strength of the field in the Rosette region. The model presented
here can be directly applied to other similar objects for which the expansion law can be derived.

\begin{acknowledgements}
We thank the referee for the useful comments. 
We acknowledge the use of the {\tt HEALPix} 
package and IRAS data. The Planck Collaboration acknowledges the
support of: ESA; CNES and CNRS/INSU-IN2P3-INP (France); ASI, CNR, and
INAF (Italy); NASA and DoE (USA); STFC and UKSA (UK); CSIC, MICINN, and
JA (Spain); Tekes, AoF, and CSC (Finland); DLR and MPG (Germany); CSA
(Canada); DTU Space (Denmark); SER/SSO (Switzerland); RCN (Norway);
SFI (Ireland); FCT/MCTES (Portugal); and DEISA (EU). A detailed
description of the Planck Collaboration and a list of its members can
be found at
\url{http://www.rssd.esa.int/index.php?project=PLANCK&page=Planck_Collaboration}. The
research leading to these results has received funding from the
European Research Council under the European Union's Seventh Framework
Programme (FP7/2007-2013)/ERC grant agreement No. 267934.
\end{acknowledgements}

\bibliographystyle{aa}
\bibliography{refs1,Planck_bib}


\appendix
\section{Magnetic field model}\label{appb}

In this appendix we present the derivation of the analytical formula that describes the
magnetic field structure in a spherical shell, following the
expansion of an ionized nebula in a uniform medium with density
$n_{0}$ and magnetic field \vec{B}$_{0}$. 

\begin{figure*}
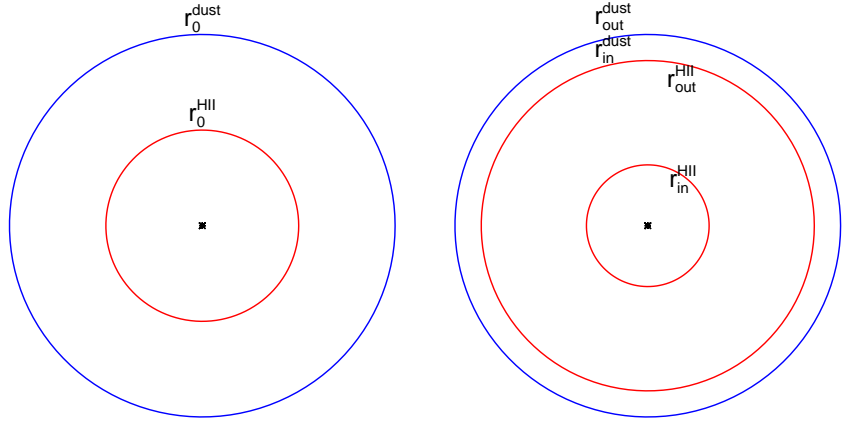

\centering
\includegraphics[scale=0.38]{Figures/figa1_a.epsi}
\includegraphics[scale=0.38]{Figures/figa1_b.epsi}
\caption{Sketch of the adopted spherical configuration. \emph{Left}:
  the initial state, with uniform density and magnetic
  field; there are no separate ionized and dust shells. \emph{Right}:
  present-day state with a cavity inside $r_{\rm 
    in}^{\hii}$, an ionized shell between $r_{\rm in}^{\hii}$ and
  $r_{\rm out}^{\hii}$, and a dust shell between $r_{\rm in}^{\rm
    dust}=r_{\rm out}^{\hii}$ and $r_{\rm out}^{\rm dust}$. The present-day radii $r_{\rm
    in}^{\hii}$, $r_{\rm out}^{\hii}$, and $r_{\rm out}^{\rm dust}$
  correspond to initial radii 0, $r_{0}^{\hii}$, and $r_{0}^{\rm
    dust}$, respectively.}
\label{fig:app0}
\end{figure*}

We start by deriving the expansion law, which will 
define how the initial uniform magnetic field is modified. 
Once the star cluster is formed, 
it ionizes the surrounding gas, which becomes overpressured and 
starts expanding at a velocity close to the ionized
gas sound speed. 
The expansion of the ionized gas, in turn, creates a cavity and sweeps
up the surrounding ISM into a thin and dense shell. The resulting
structure, as observed at the present time, is composed of a cavity of
radius $ r_{\rm in}^{\hii}$, surrounded by a thick shell of ionized
gas extending from $r_{\rm in}^{\hii}$ to $r_{\rm out}^{\hii}$, itself
surrounded by a thin dust shell extending from $r_{\rm in}^{\rm dust}$
to $r_{\rm out}^{\rm dust}$ (see Fig. \ref{fig:app0}). 
For the Rosette, $r_{\rm in}^{\hii}=7$\,pc and $r_{\rm
  out}^{\hii}=19$\,pc, as derived from the radial distribution of the
radio emission (Sect. \ref{sec:polrm}); $r_{\rm in}^{\rm dust}=18$\,pc
and $r_{\rm out}^{\rm dust}=22$\,pc, measured 
from the \Planck\ 353\,GHz latitude cut through the centre of the
shell (Sect. \ref{sec:int}). Thus, the
inner radius of the dust shell is slightly smaller than the outer
radius of the \hii~region. This is not surprising, as the Rosette is
an ionization bounded Str{\"o}mgren sphere and thus the two shells
are expected to overlap at the boundary, where the ionized and
neutral gases are mixed. However, for the sake of simplicity, we take
$r_{\rm in}^{\rm dust}= r_{\rm out}^{\hii} =19$\,pc.

We use a parameter grid of 1\,pc resolution for all the
  radii. Therefore, we adopt an uncertainty of 1\,pc in $ r_{\rm in}^{\hii}$ and
  $r_{\rm out}^{\hii}$. We ascribe a larger uncertainty of 2\,pc to
$ r_{\rm in}^{\rm dust}$ and $ r_{\rm out}^{\rm dust}$; this reflects the variation of the fitted radii when
considering the longitude or radial profile of the dust emission,
which are affected by the presence of the Mon OB2 cloud. 

\begin{figure}
\centering
\hspace{-0.5cm}
\includegraphics[scale=0.55]{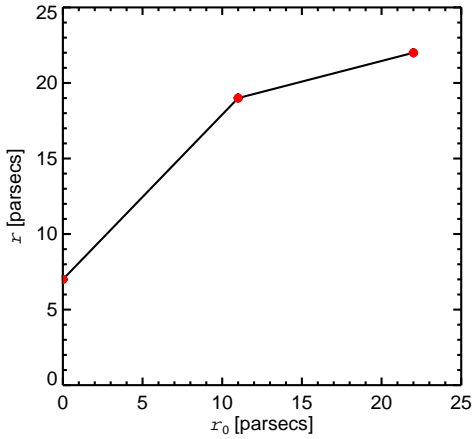}
\caption{Radial expansion law derived from the three known radii in
 the initial, $r_{0}$, and final, $r$, states (red filled
   circles). The black curve
 is described by a piecewise linear function of the form $r=\alpha r_{0}+\beta$,
 with $\alpha=1.09$, $\beta=7$\,pc for $r _{0}< r_{0}^{\hii}$
and $\alpha=0.27$, $\beta=16$\,pc for $r_{0}^{\hii}
\leq r_{0} \leq r_{0}^{\rm dust}$.}
\label{fig:app3}
\end{figure}

We now denote by $r_{0}$ the initial radius of a particle currently at
radius $r$. In our simplified model, the initial radius of a particle
currently at $r_{\rm in}^{\hii}$ is simply $r_{0}=0$, while the
initial radii of particles currently at $r_{\rm out}^{\hii}$ and
$r_{\rm out}^{\rm dust}$ can be denoted by $r_{0}^{\hii}$ and
$r_{0}^{\rm dust}$, respectively. We emphasize that $r_{0}^{\hii}$ and
$r_{0}^{\rm dust}$ are just two working quantities, which do not
correspond to any physical boundaries. The value of $r_{0}^{\rm dust}$
can be obtained by noting that a particle currently at $r_{\rm
  out}^{\rm dust}$ has just been reached by the expanding shell and
has not yet moved from its initial radius $r_{0}^{\rm dust}$, so that
$r_{0}^{\rm dust}=r_{\rm out}^{\rm dust}=22$\,pc. The value of
$r_{0}^{\hii}$ can be inferred from the conservation of mass. In the initial state
\begin{align}
\label{model0}
& \frac{4}{3}\pi (r_{0}^{\hii})^{3} n_{0} = M_{\hii}, \\
& \frac{4}{3}\pi \left( (r_{0}^{\rm dust})^{3} - (r_{0}^{\hii})^{3}\right) n_{0} = M_{\rm dust}, 
\end{align}
where $M_{\hii}$ is the mass inside the sphere with radius
$r_{0}^{\hii}$ and $M_{\rm dust}$ is the mass inside the shell with
inner and outer radii $r_{0}^{\hii}$ and $r_{0}^{\rm dust}$,
respectively, as shown in the left panel of Fig. \ref{fig:app0}. 
From the previous equations we can write
\begin{equation}\label{model1}
r_{0}^{\hii} = r_{0}^{\rm dust} \left(\frac{M_{\rm dust}}{M_{\hii}} +1\right)^{-1/3}.
\end{equation}
We calculate the mass of ionized gas based on the size of the nebula and on
the mean electron density of 12.3\,cm$^{-3}$, derived from the radio
data (Sect. \ref{sec:polrm}). Taking into account the
contribution from ionized helium \citep{Celnik:1985}, we
obtain $M_{\hii} =1.2\times10^{4}$\,M$_{\odot}$. For the mass of the
dust shell we use the results of \citet{Heyer:2006} (Sect. \ref{sec:intro2}), 
$M_{\rm dust} =8.6\times10^{4}$\,M$_{\odot}$. As a result, 
$r_{0}^{\hii}=11$\,pc. With the three known radii in the initial
state, $(0, r_{0}^{\hii}, r_{0}^{\rm dust})$, and their
corresponding values in the present state, $(r_{\rm in}^{\hii}, r_{\rm
out}^{\hii}, r_{\rm out}^{\rm dust})$, we can derive an expansion law
as shown in Fig. \ref{fig:app3}. With only three data points,
  we consider the simplest description of the expansion by writing $r$
as a piecewise linear function of $r_{0}$. This is in any case sufficient to
describe the two clear regimes seen in Fig. \ref{fig:app3}: the
expansion of the \hii~region (slope larger than 1) and the compression of the ISM within the
outer shell (slope smaller than 1). The expansion law is hence given by
$r=\alpha r_{0}+\beta$, where $\alpha=1.09$, $\beta=7$\,pc for the
\hii~region ($r_{0} < r_{0}^{\hii}$)
and $\alpha=0.27$, $\beta=16$\,pc for the dust shell ($r_{0}^{\hii}
\leq r_{0} \leq r_{0}^{\rm dust}$).

Now consider a cartesian coordinate system with $z$-axis along the
line of sight and $x$- and $y$-axes in the plane of the sky, along the
trace of the Galactic plane and along the rotation axis,
respectively. The initial uniform magnetic field can be written in
these cartesian coordinates as
\begin{equation}\label{model2}
\vec{B}_{0} = B_{0x}\, \vec{e}_{x} +  B_{0y}\, \vec{e}_{y} + B_{0z}\, \vec{e}_{z},
\end{equation}
with
\begin{align}
\label{model2a}
& B_{0x} = B_{0} \sin\theta_{0}\cos\phi_{0} \nonumber \\
& B_{0y} = B_{0}\sin\theta_{0}\sin\phi_{0}   \nonumber \\
& B_{0z} =  B_{0} \cos\theta_{0},
\end{align}
where $\theta_{0}$ and $\phi_{0}$ are the
polar and azimuthal angles, respectively, i.e. $\theta_{0}$ is the
angle between \vec{B}$_{0}$ and the line of sight, while $\phi_{0}$
gives the plane-of-the-sky direction of \vec{B}$_{0}$ with respect to
the Galactic plane ($\phi_{0}=0\degr$ for \vec{B}$_{0}$ parallel to the
Galactic plane).

The vector potential associated with \vec{B}$_{0}$ is
\begin{equation}\label{model3}
\vec{A}_{0} = B_{0y}z\,\vec{e}_{x} + B_{0z}x\,\vec{e}_{y}+B_{0x}y\,\vec{e}_{z},
\end{equation}
which satisfies the condition $\vec{B}_{0}=\nabla \times
\vec{A}_{0}$. 
We use the frozen-in approximation and assume that the magnetic field
evolves from the initially uniform configuration following the
radial expansion of the gas. The vector potential in the final state is given by 
(see equations 4 to 10 in \citealt{Parker:1970})
\begin{equation}\label{model4}
\vec{A(\vec{r})} = (\nabla \vec{r}_{0})\cdot
\vec{A}_{0}(\vec{r}_{0}),
\end{equation}
and the resulting magnetic field works out to be 
\begin{align}
\label{model7}
\vec{B(\vec{r})} = \left(\frac{r_{0}}{r}\right)^{2}B_{0r}\,\vec{e}_{r}+\frac{r_{0}}{r}\frac{dr_{0}}{dr}
(B_{0\theta}\,\vec{e}_{\theta} + B_{0\phi}\,\vec{e}_{\phi}). 
\end{align}
The previous equation, written in spherical coordinates, clearly shows
the change in both the normal (radial) and tangential components of
the magnetic field, relative to the expansion front.

\begin{figure}
\centering
\hspace{-0.5cm}
\includegraphics[scale=0.45]{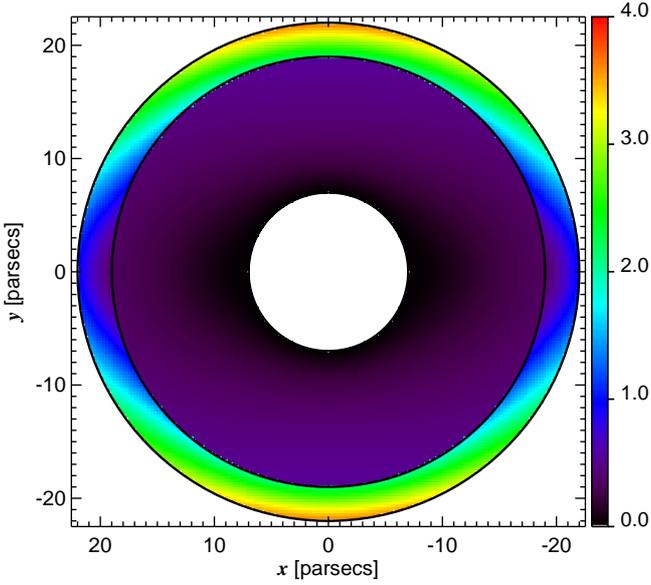}
\caption{Ratio of the field strength $B/B_{0}$ for a vertical slice through
  the centre of the bubble-shell structure ($xy$ plane). The initial field
  \vec{B}$_{0}$ is on the plane of the sky ($\theta_{0}=90\degr$) and
  along $x$ ($\phi_{0}=0\degr$). The three black circles delineate the radii
of the Rosette \hii\ and dust shells:  $r_{\rm in}^{\hii}=7$\,pc; $r_{\rm
  out}^{\hii}=r_{\rm in}^{\rm dust}=19$\,pc; and $r_{\rm out}^{\rm dust}=22$\,pc.}
\label{fig:app4}
\end{figure}

We create a 3D cartesian grid of $181^{3}=5\,929\,741$ voxels, each equivalent to
0.25\,pc ($0\farcm5$ at the distance of the Rosette), with the
bubble-shell structure located at the origin. The resolution of the
model is finer than that of the observations, which is needed to
have the required sampling to compute the integrals along the line of sight.
We use Eq. (\ref{model7}) along with the expansion law of Fig. \ref{fig:app3} to calculate the magnetic
field strength in every pixel of the grid. Figure \ref{fig:app4} shows
how the field strength in the shell, $B$, varies relative to the 
initial field strength, $B_{0}$. The map corresponds to a vertical cut through
the centre of the shell for an initial field with
$(\theta_{0},\phi_{0})=(90\degr,0\degr)$, therefore on the plane of
the sky and along the Galactic plane. Figure \ref{fig:app4}
illustrates that the largest compression of the field lines occurs towards the
equator of the shell, or in the direction perpendicular to the initial
field \vec{B}$_{0}$, where the ratio $B/B_{0}$ is seen to increase from
the centre to the outer radius of the dust shell. The change
  in expansion law at the boundary between the
\hii\ and dust shells, $r=19$\,pc, results in a discontinuity in the
tangential component of the magnetic field.
On the other hand,
close to the poles of the shell, or along \vec{B}$_{0}$, the field
lines are little disturbed, with the ratio $B/B_{0}$
  continuously increasing from
the centre to $B/B_{0}=1$ at the boundary of the dust
shell.
Owing to the axial symmetry of the
magnetic field model, the map of Fig. \ref{fig:app4} is reproduced in
every plane about the direction of the initial field \vec{B}$_{0}$.

\end{document}